\PassOptionsToPackage{dvipsnames}{xcolor}
\documentclass[year=26]{fmcad}

\usepackage{amsmath}
\usepackage{mathtools}
\usepackage{stmaryrd}
\usepackage{amssymb}
\usepackage{wasysym}
\usepackage{graphicx}
\usepackage{subcaption}
\usepackage{caption}
\usepackage{booktabs}
\usepackage{listings}
\usepackage{tabularx}

\usepackage{tikz}
\usetikzlibrary{calc}
\usetikzlibrary{shapes.geometric, shapes.arrows}
\usetikzlibrary{arrows.meta,arrows, positioning, fit, backgrounds}
\usetikzlibrary{shapes.multipart, arrows.meta, positioning}
\usepackage{xkvltxp}
\usepackage[status=draft]{fixme}
\fxsetup{
  layout=inline,
  inlineface=\itshape,
  theme=color
}

\usepackage{xspace}

\newcommand{\true}[0]{{\mathtt{true}}}
\newcommand{\false}[0]{{\mathtt{false}}}

\newcommand{\PastLTL}[0]{\textsf{PastLTL}\xspace}
\newcommand{\PastMTL}[0]{\textsf{PastMTL}\xspace}

\DeclareMathOperator{\since}{\texttt{S}}
\DeclareMathOperator{\once}{\texttt{P}}
\DeclareMathOperator{\hist}{\texttt{H}}
\DeclareMathOperator*{\prev}{\texttt{Y}}

\newcommand{\xt}[2]{\texttt{[{\tiny#1:#2}]}}

\newcommand{\histx}[2]{\hist_{\xt{#1}{#2}}}

\newcommand{\sincex}[2]{\since_{\xt{#1}{#2}}}

\begin{document}

\title{Multi-Property Temporal Logic Monitoring}

\author{
    \IEEEauthorblockN{Arınç Demir}
    \IEEEauthorblockA{
    \textit{Boğaziçi University}\\
    Istanbul, Türkiye \\
    arinc.demir@std.bogazici.edu.tr}
    \and
    \IEEEauthorblockN{Dogan Ulus}
    \IEEEauthorblockA{
    \textit{Boğaziçi University}\\
    Istanbul, Türkiye \\
    dogan.ulus@bogazici.edu.tr}
}

\maketitle

\begin{abstract}
Runtime verification enables checking temporal logic specifications over individual execution traces and offers a scalable alternative to exhaustive formal verification.
In practice, systems must satisfy dozens to hundreds of temporal properties simultaneously; however, existing approaches monitor each property in isolation, resulting in redundant computation and limited scalability.
In this work, we present an online multi-property monitoring framework that compiles past-time LTL and MTL specifications into a shared directed acyclic graph of subformulas with one output per property.
Unlike prior approaches that construct monitors independently, our method extends compositional sequential network-based temporal logic monitor construction to a shared setting, enabling reuse of intermediate results across properties while preserving their individual structure.
Central to our approach is a data-oriented execution model based on an arena-allocated, double-buffered layout that stores intermediate results for each subformula in compact, contiguous memory. 
This design favors spatial locality and enables incremental updates with minimal overhead.
Experimental results demonstrate per-property throughput improvements of $2\times$ to $4.5\times$ and $6\times$ to $12\times$ in multi-property configurations compared to conventional single-property monitoring, enabling scalability to large specification sets and deployment in high-performance and resource-constrained systems.
\end{abstract}

\section{Introduction}
\label{sec:introduction}

Runtime verification has emerged as a practical technique for ensuring the correctness of complex systems by checking formal specifications against execution traces at runtime. Unlike exhaustive verification methods such as model checking or theorem proving, runtime verification operates on concrete executions, making it particularly suitable for systems where full state-space exploration is infeasible due to complexity, concurrency, or interaction with unpredictable environments. Among specification formalisms, Linear Temporal Logic (LTL)~\cite{pnueli1977ltl} and Metric Temporal Logic (MTL)~\cite{koymans1990mtl} are particularly expressive, enabling the capture of qualitative temporal relations and quantitative timing constraints that are essential in domains such as embedded systems, cyber-physical systems, and real-time software. 

In many realistic scenarios, systems must be verified against not just a single property, but a collection of many small temporal logic specifications simultaneously. A straightforward approach instantiates an independent monitor for each property, each processing the same execution trace. However, properties often share common subformulas, resulting in repeated evaluation of identical substructures across monitors, while intermediate results are maintained in isolation. As the number and complexity of monitored properties increase, this redundancy leads to poor performance and limited scalability.

\begin{figure}[t]
\centering
\begin{subfigure}[t]{\columnwidth}
\centering
\resizebox{0.9\columnwidth}{!}{%
\begin{tikzpicture}[
    >=Latex,
    every node/.style={draw, circle, thick, font=\bfseries\scriptsize, inner sep=0pt},
    unary/.style={minimum size=5mm},
    binary/.style={minimum size=5mm},
    edge/.style={->, thick}
]

\node[unary,  fill=red!35]      (n1)  at (0.0,  0.98) {1};
\node[unary,  fill=blue!35]     (n2)  at (0.2,  0.00) {2};
\node[unary,  fill=green!35]    (n3)  at (-0.1, -0.84) {3};

\node[binary, fill=orange!40]   (n4)  at (1.4,  0.56) {4};
\node[binary, fill=yellow!45]   (n5)  at (1.3, -0.49) {5};

\node[unary,  fill=purple!35]   (n6)  at (2.7,  0.91) {6};
\node[binary, fill=cyan!35]     (n7)  at (2.9,  0.00) {7};
\node[unary,  fill=pink!40]     (n8)  at (2.5, -0.77) {8};

\node[binary, fill=teal!35]     (n9)  at (4.1,  0.49) {9};
\node[binary, fill=lime!40]     (n10) at (4.0, -0.42) {10};

\node[unary, double, fill=brown!35]    (n11) at (5.3,  0.63) {11};
\node[unary, double, fill=magenta!35]  (n12) at (5.2, -0.56) {12};

\draw[edge] (n1) -- (n4);
\draw[edge] (n2) -- (n4);
\draw[edge] (n2) -- (n5);
\draw[edge] (n3) -- (n5);

\draw[edge] (n4) -- (n6);
\draw[edge] (n4) -- (n7);
\draw[edge] (n5) -- (n7);
\draw[edge] (n5) -- (n8);

\draw[edge] (n6) -- (n9);
\draw[edge] (n7) -- (n9);
\draw[edge] (n7) -- (n10);
\draw[edge] (n8) -- (n10);

\draw[edge] (n9) -- (n11);
\draw[edge] (n10) -- (n12);

\end{tikzpicture}
}
\caption{}
\label{fig:computation-graph}
\end{subfigure}

\vspace{0.2em}

\begin{subfigure}[t]{\columnwidth}
\centering
\begin{tikzpicture}

\def\cell{0.25}
\def\cols{20}
\def\rows{5}

\foreach \start/\stop/\clr in {
    0/17/red!30,
    18/36/blue!30,
    37/51/green!30,
    52/78/orange!30,
    79/89/purple!30
}{
    \foreach \i in {\start,...,\stop} {
        \pgfmathtruncatemacro{\col}{mod(\i,\cols)}
        \pgfmathtruncatemacro{\row}{int(\i/\cols)}
        \fill[\clr] (\col*\cell, 1.0-\row*\cell) rectangle ++(\cell,\cell);
    }
}

\foreach \start/\stop/\clr in {
    0/9/red!30,
    10/33/blue!30,
    34/58/green!30,
    59/71/orange!30,
    72/78/purple!30
}{
    \foreach \i in {\start,...,\stop} {
        \pgfmathtruncatemacro{\col}{mod(\i,\cols)}
        \pgfmathtruncatemacro{\row}{int(\i/\cols)}
        \fill[\clr] (\col*\cell, 2.5-\row*\cell) rectangle ++(\cell,\cell);
    }
}

\draw[step=\cell, thick, black] (0,0) grid (5,1.25);       
\draw[step=\cell, thick, black] (0,1.5) grid (5,2.75); 
\draw[thick, black] (0,1.5) -- (5,1.5);

\node[anchor=east] (prevnode) at (-0.1,0.6) {};
\node[anchor=east] (currnode) at (-0.1,2.1) {};

\draw[<->, very thick] (prevnode) to[out=160, in=200] (currnode);

\node[anchor=west] at (5.1,0.6) {\small $B_{\mathrm{prev}}$};
\node[anchor=west] at (5.1,2.1) {\small $B_{\mathrm{cur}}$};

\end{tikzpicture}
\caption{}
\label{fig:buffer-transition}
\end{subfigure}
\caption{Overview of multi-property monitoring framework.\newline (a) A unified computation graph built from multiple properties.\newline (b) A double buffering implementation for storing and updating internal node states efficiently at runtime.}
\label{fig:graph-buffer-overview}
\end{figure}
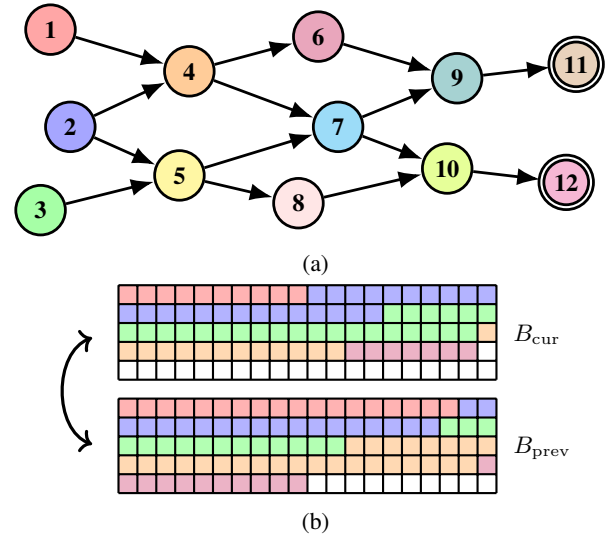

In this work, we present an online monitoring framework, \emph{LoomRV}, that compiles a collection of past-time LTL (\PastLTL) and MTL (\PastMTL) specifications into a single unified monitor with one output per property, supporting both discrete and dense time models. Our approach extends compositional sequential network-based monitor synthesis~\cite{ulus2026online} by merging equivalent subformulas across specifications into a shared computation structure. In the original construction, a temporal logic specification is translated into a sequential network in which each subformula is represented by a node with an associated state variable. The network follows the syntactic structure of the formula, and each node computes its value from the current input and its dependencies. Temporal operators are realized through stateful update rules that propagate information from their previous valuations.

Figure~\ref{fig:graph-buffer-overview} illustrates the two orthogonal aspects of our proposed multi-property extension: a unified computation graph and a data-oriented execution model.
The unified computation graph, shown in Figure~\ref{fig:computation-graph}, is compiled from multiple specifications into a single shared structure.
Each node represents a unique subformula, and directed edges encode the data dependencies governing the propagation of intermediate values during evaluation. 
Equivalent subformulas across specifications are merged into a single node, allowing results to be computed once and reused. 
The graph can therefore be evaluated in a single pass over the input trace, producing verdicts for all properties simultaneously.

The second aspect addresses runtime efficiency through a data-oriented execution model.
Eliminating common subformulas reduces redundant computation; however, the increased number of nodes and irregular memory access patterns can still degrade performance. 
To mitigate this, we store all subformula states in compact, contiguous memory buffers as shown in Figure~\ref{fig:buffer-transition}. 
The monitor alternates between two buffers representing successive time steps, since each state depends only on its previous value and the current valuations. 
This double-buffered layout separates read and write phases, enabling structured, sequential updates without overwriting required data. 
Together, the shared graph and data-oriented execution model enable efficient single-pass monitoring of multiple specifications.

\begin{table}[t]
\centering
\caption{Comparison of multi-property monitoring}
\label{tab:single-vs-multi}
\begin{tabularx}{\columnwidth}{@{}lXX@{}}
\toprule
& Single-property\newline monitoring~\cite{ulus2026online} 
& Multi-property\newline monitoring (This work) \\
\midrule
Specification 
& Single property 
& Multiple properties \\

Monitor structure
& One dedicated monitor
& One shared monitor \\

Construction 
& Independent
& Unified \\

Shared subformulas
& Duplicated 
& Unique \\

Computation topology 
& Tree 
& DAG \\

Computation reuse 
& None 
& Shared \\

Update semantics 
& Synchronous 
& Synchronous \\

Evaluation 
& Multiple passes 
& Single pass \\

Memory layout 
& Scattered memory allocations on heap
& Double-buffered contiguous storage \\

Outputs 
& One per monitor 
& One per property \\

\bottomrule
\end{tabularx}
\end{table}

We summarize the key architectural and execution-level differences between baseline single-property monitoring and our unified framework in Table~\ref{tab:single-vs-multi}. The comparison highlights the transition from independent, tree-structured monitors with duplicated subformulas and scattered state storage to a shared DAG-based representation with explicit subformula reuse and a data-oriented, double-buffered execution model.

The remainder of the paper is organized as follows. Section~\ref{sec:background} introduces the necessary preliminaries on \PastMTL\ and briefly reviews the monitoring approach underlying existing tools, which serves as the baseline for our work. Section~\ref{sec:methods} presents our multi-property monitoring framework, detailing how specifications are compiled and deduplicated into a shared computation graph, followed by the data-oriented execution model that drives the evaluation. Section~\ref{sec:evaluation} describes the implementation and experimental setup, and evaluates the proposed approach against existing techniques in terms of performance and scalability. Section~\ref{sec:related} discusses prior work in runtime verification and related areas. Section~\ref{sec:conclusion} discusses our contributions and outlines directions for future work.

\section{Background}
\label{sec:background}
Past-time Linear Temporal Logic (\PastLTL) extends propositional logic with past temporal operators, while past-time Metric Temporal Logic (\PastMTL) further augments these operators with timing constraints.
This section reviews the construction of sequential network-based monitors for such temporal logic specifications, following the approach introduced in~\cite{ulus2026online}.
For clarity of exposition, we focus on discrete-time \PastMTL. 
The underlying execution model and compositional monitor construction extend to dense-time settings in an analogous manner, although semantic differences between the two settings require additional considerations.

Given a finite set $P$ of atomic predicates, the formulas of discrete-time \PastMTL\ are inductively defined by the grammar:
\begin{equation*}
\varphi\ \coloneqq\ \bot\ |\ p \ |\ \neg \varphi\ |\ \varphi_{1} \wedge \varphi_{2}\ |\ \prev\varphi \ |\ \varphi_{1} \sincex{a}{b} \varphi_{2}
\end{equation*}
where $p \in P$, the unary operator $\prev$ denotes the \emph{Previous} modality, and the binary operator $\sincex{a}{b}$ denotes the timed version of the \emph{Since} temporal modality, whose range of quantification is restricted by interval bounds $a$ and $b$. 
The satisfaction relation of a \PastMTL\ formula $\varphi$ over a discrete-time trace $w$ at time $t$, denoted as $(w, t) \vDash \varphi$, is defined inductively as follows:
\begin{equation*}
\begin{array}{rlcl}
(w,t)&\vDash \bot & \ \leftrightarrow\ & \false\\
(w,t)&\vDash p & \ \leftrightarrow\ & w_p(t) = \true  \\
(w,t)&\vDash \neg \varphi & \ \leftrightarrow\ & (w,t) \nvDash \varphi \\
(w,t)&\vDash \varphi_{1} \wedge \varphi_{2} & \ \leftrightarrow\ & (w,t) \vDash \varphi_{1} \text{ and } (w, t) \vDash \varphi_{2}\\
(w,t)&\vDash \prev\varphi & \ \leftrightarrow\ & (w,t - 1) \vDash \varphi\\
(w,t)&\vDash \varphi_{1}\sincex{a}{b} \varphi_{2} & \ 
 \leftrightarrow\ & \exists t' < t.\ (w, t') \vDash \varphi_{2} \text{ and }\\
 & & & \forall t'' \in (t', t).\ (w, t'') \vDash \varphi_{1}\text{ and }\\
 & & & \quad\quad t-b \leq t' < t-a\\
\end{array}
\label{DefProp}
\end{equation*}

The remaining Boolean operators such as disjunction ($\vee$) and logical implication ($\rightarrow$) are defined in the standard way. 
From the timed Since $\sincex{a}{b}$ modality, one can derive the \emph{Past Eventually} ($\once$) and \emph{Past Always} ($\hist$) temporal modalities, along with their timed variants. 
A temporal operator is called \emph{untimed} if it is not parameterized by interval bounds.

The monitor construction for a single formula proceeds by associating a computation node with each subformula, together with an update rule determined by its syntactic type.
At runtime, the state of each node is stored explicitly and updated at every time step, making the memory access pattern a central aspect of the monitoring procedure.
Atomic predicates and Boolean operators require only a single Boolean value representing the current valuation. 
In contrast, untimed past temporal operators such as Previous and Since depend on history and therefore require access to the state at the previous step. 
For metric temporal operators, the node must additionally maintain a symbolic set of marked time points determined by the update rule. 
This set can be represented compactly using intervals, with its size depending on the input trace and bounded by the interval parameters of the operator. 
In practice, worst-case growth is rarely observed on typical traces.

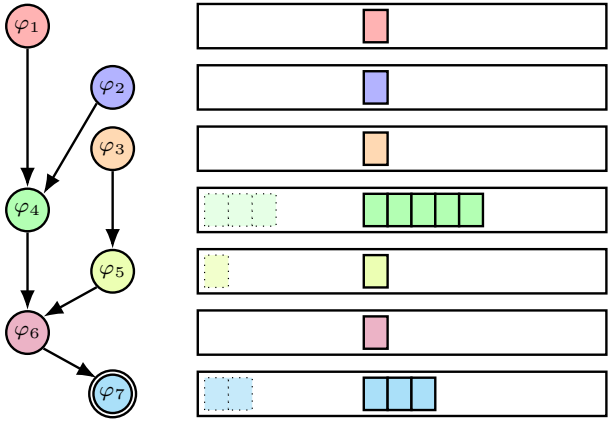
\begin{figure}[t]
\centering
\resizebox{0.9\columnwidth}{!}{%
\begin{tikzpicture}[
    >=Latex,
    every node/.style={font=\bfseries\scriptsize},
    outer/.style={draw, thick}
]

\def\xnode{0.0}
\def\xbuf{1.0}
\def\W{4.8}
\def\H{0.52}
\def\cw{0.28}
\def\dy{0.72}

\node(phi1)[draw, circle, thick, fill=red!30, minimum size=5mm, inner sep=0pt] at (\xnode-1,0) {$\varphi_{1}$};
\draw[outer] (\xbuf,0-\H/2) rectangle ++(\W,\H);
\draw[fill=red!30, thick] (\xbuf+1.95,0-0.19) rectangle ++(\cw,0.38);

\node(phi2)[draw, circle, thick, fill=blue!30, minimum size=5mm, inner sep=0pt] at (\xnode,-\dy) {$\varphi_{2}$};
\draw[outer] (\xbuf,-\dy-\H/2) rectangle ++(\W,\H);
\draw[fill=blue!30, thick] (\xbuf+1.95,-\dy-0.19) rectangle ++(\cw,0.38);

\node(phi4)[draw, circle, thick, fill=orange!30, minimum size=5mm, inner sep=0pt] at (\xnode,-2*\dy) {$\varphi_{3}$};
\draw[outer] (\xbuf,-2*\dy-\H/2) rectangle ++(\W,\H);
\draw[fill=orange!30, thick] (\xbuf+1.95,-2*\dy-0.19) rectangle ++(\cw,0.38);

\node(phi3)[draw, circle, thick, fill=green!30, minimum size=5mm, inner sep=0pt] at (\xnode-1,-3*\dy) {$\varphi_{4}$};
\draw[outer] (\xbuf,-3*\dy-\H/2) rectangle ++(\W,\H);
\foreach \i in {0,1,2}{
    \draw[fill=green!10, dotted] (\xbuf+0.08+\i*\cw,-3*\dy-0.19) rectangle ++(\cw,0.38);
}
\foreach \i in {0,1,2,3,4}{
    \draw[fill=green!30, thick] (\xbuf+1.95+\i*\cw,-3*\dy-0.19) rectangle ++(\cw,0.38);
}

\node(phi5)[draw, circle, thick, fill=lime!30, minimum size=5mm, inner sep=0pt] at (\xnode,-4*\dy) {$\varphi_{5}$};
\draw[outer] (\xbuf,-4*\dy-\H/2) rectangle ++(\W,\H);
\draw[fill=lime!20, dotted] (\xbuf+0.08,-4*\dy-0.19) rectangle ++(\cw,0.38);
\draw[fill=lime!30, thick] (\xbuf+1.95,-4*\dy-0.19) rectangle ++(\cw,0.38);

\node(phi6)[draw, circle, thick, fill=purple!30, minimum size=5mm, inner sep=0pt] at (\xnode-1,-5*\dy) {$\varphi_{6}$};
\draw[outer] (\xbuf,-5*\dy-\H/2) rectangle ++(\W,\H);
\draw[fill=purple!30, thick] (\xbuf+1.95,-5*\dy-0.19) rectangle ++(\cw,0.38);

\node(phi7)[draw, circle, double, thick, fill=cyan!30, minimum size=5mm, inner sep=0pt] at (\xnode,-6*\dy) {$\varphi_{7}$};
\draw[outer] (\xbuf,-6*\dy-\H/2) rectangle ++(\W,\H);
\foreach \i in {0,1}{
    \draw[fill=cyan!20, dotted] (\xbuf+0.08+\i*\cw,-6*\dy-0.19) rectangle ++(\cw,0.38);
}
\foreach \i in {0,1,2}{
    \draw[fill=cyan!30, thick] (\xbuf+1.95+\i*\cw,-6*\dy-0.19) rectangle ++(\cw,0.38);
}

\draw[->, thick] (phi1.south) -- (phi3.north);
\draw[->, thick] (phi2.south west) -- (phi3.north east);
\draw[->, thick] (phi3.south) -- (phi6.north);
\draw[->, thick] (phi4.south) -- (phi5.north);
\draw[->, thick] (phi5.south west) -- (phi6.north east);
\draw[->, thick] (phi6.south east) -- (phi7.north west);
\end{tikzpicture}%
}
\caption{Object-oriented execution model where each computation node maintains its own local storage in memory.}
\label{fig:object-oriented}
\end{figure}

We illustrate the baseline monitor construction using a representative example that demonstrates how different operator types are translated into a computation graph. Consider the \PastMTL formula
\[
\varphi \coloneqq \histx{a}{b}(\prev(r) \wedge (p\ \sincex{c}{d}\ q)),
\]
which has seven subformulas, including $\varphi$ itself. The construction associates one computation node with each subformula. For this formula, the resulting computation graph contains three atomic-predicate nodes for $\varphi_{1}\coloneqq p$, $\varphi_{2}\coloneqq q$, and $\varphi_{3}\coloneqq r$, one untimed temporal node for $\varphi_{5}\coloneqq\prev(\varphi_{3})$, one Boolean node for $\varphi_{6}\coloneqq \varphi_{4} \wedge \varphi_{5}$,  and two metric temporal nodes for $\varphi_{4}\coloneqq \varphi_{1}\sincex{c}{d}\varphi_{2}$ and $\varphi=\varphi_{7}\coloneqq\histx{a}{b}\varphi_{6}$.


Figure~\ref{fig:object-oriented} illustrates the resulting computation graph together with its memory organization. The nodes are shown on the left, with $\varphi_{7}$ representing the top-level formula $\varphi$, and the storage associated with each node is depicted on the right. In this execution model, each computation node maintains its own local state in memory, resulting in a per-node, object-oriented storage layout. Predicate, Boolean, and untimed temporal nodes require only a constant amount of storage, since their evaluation depends on the current input and, at most, the immediately preceding state. In contrast, metric temporal nodes maintain sets of intervals with variable sizes.

A crucial observation is that memory is fragmented across nodes: each subformula allocates and manages its own buffer independently. As a result, the overall memory layout consists of multiple disjoint regions, with no guarantee of spatial locality between dependent nodes. During execution, updating a node requires accessing its own buffer as well as the buffers of its predecessors, leading to scattered memory accesses. This object-oriented layout directly mirrors the compositional structure of the formula, but it introduces overhead due to redundant allocation, fragmented memory access, and lack of coordination between node states. These limitations motivate the transition to a unified, data-oriented memory organization discussed later.

\section{Methods}
\label{sec:methods}

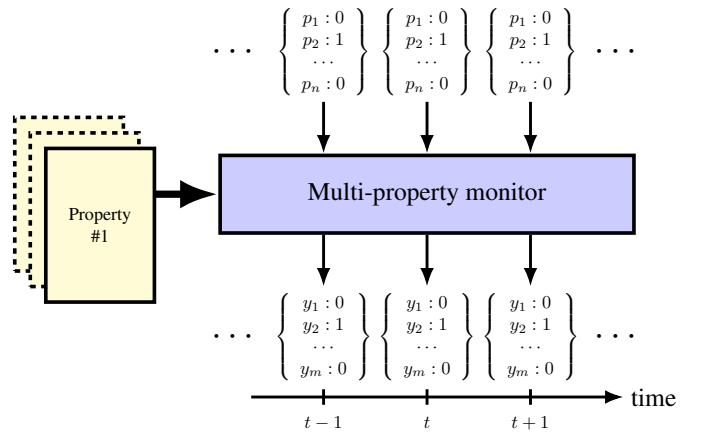
\begin{figure}[b]
\centering
\resizebox{\columnwidth}{!}{%
\begin{tikzpicture}[
    >=Latex,
    box/.style={draw, fill=blue!20, minimum width=8cm, minimum height=1.5cm, align=center, line width=2pt},
    smallbox/.style={draw, fill=yellow!20, minimum width=2.10cm, minimum height=2.97cm, align=center, line width=2pt},
    vec/.style={align=center},
    arr/.style={-Latex, ultra thick},
    compilearrow/.style={draw, fill=blue!20, single arrow, single arrow head extend=2mm,
                         minimum height=1cm, minimum width=1.6cm}
]

\node[smallbox, dashed] (x1front) at (0,0) {\Large X1};
\node[box, right=1.8cm of x1front] (monitor) {\Large Multi-property monitor};
\draw[arr, line width=4pt]  (x1front.east) -- (monitor.west);

\node[smallbox, dashed, behind path, xshift=0.3cm, yshift=-0.3cm] at (x1front.center) {};
\node[smallbox, behind path, xshift=0.6cm, yshift=-0.6cm] at (x1front.center) {Property\\\#1};


\node[vec, above=1cm of monitor, xshift=-2cm] (p1) {$\left\{\begin{array}{c}p_1: 0\\p_2: 1\\\cdots\\p_n: 0\end{array}\right\}$};
\node[vec, above=1cm of monitor] (p2) {$\left\{\begin{array}{c}p_1: 0\\p_2: 1\\\cdots\\p_n: 0\end{array}\right\}$};
\node[vec, above=1cm of monitor, xshift=2cm] (p3) {$\left\{\begin{array}{c}p_1: 0\\p_2: 1\\\cdots\\p_n: 0\end{array}\right\}$};

\node[vec, below=1cm of monitor, xshift=-2cm] (y1) {$\left\{\begin{array}{c}y_1: 0\\y_2: 1\\\cdots\\y_m: 0\end{array}\right\}$};
\node[vec, below=1cm of monitor] (y2) {$\left\{\begin{array}{c}y_1: 0\\y_2: 1\\\cdots\\y_m: 0\end{array}\right\}$};
\node[vec, below=1cm of monitor, xshift=2cm] (y3) {$\left\{\begin{array}{c}y_1: 0\\y_2: 1\\\cdots\\y_m: 0\end{array}\right\}$};

\node[scale=2] at ($(p1)-(1.7,0)$) {$\cdots$};
\node[scale=2] at ($(p3)+(1.7,0)$) {$\cdots$};
\node[scale=2] at ($(y1)-(1.7,0)$) {$\cdots$};
\node[scale=2] at ($(y3)+(1.7,0)$) {$\cdots$};

\draw[arr] (p1.south) -- ++(0,-0.6) -- ($(monitor.north)+(-2cm,0)$);
\draw[arr] (p2.south) -- ++(0,-0.6) -- (monitor.north);
\draw[arr] (p3.south) -- ++(0,-0.6) -- ($(monitor.north)+(+2cm,0)$);

\draw[arr] ($(monitor.south)+(-2cm,0)$) -- ++(0,-0.6) -- (y1.north);
\draw[arr] (monitor.south) -- ++(0,-0.6) -- (y2.north);
\draw[arr] ($(monitor.south)+(+2cm,0)$) -- ++(0,-0.6) -- (y3.north);

\draw[-Latex, ultra thick] ($(y1.south)+(-1.4cm,-0.2cm)$) -- ($(y3.south)+(1.8cm,-0.2cm)$)
    node[right] {\Large time};

\draw[ultra thick] ($(y1.south)+(0,-0.05cm)$) -- ($(y1.south)+(0,-0.35cm)$);
\draw[ultra thick] ($(y2.south)+(0,-0.05cm)$) -- ($(y2.south)+(0,-0.35cm)$);
\draw[ultra thick] ($(y3.south)+(0,-0.05cm)$) -- ($(y3.south)+(0,-0.35cm)$);

\node[below=3pt] at ($(y1.south)+(0,-0.35cm)$) {$t-1$};
\node[below=3pt] at ($(y2.south)+(0,-0.35cm)$) {$t$};
\node[below=3pt] at ($(y3.south)+(0,-0.35cm)$) {$t+1$};

\end{tikzpicture}%
}
\caption{Multi-property monitor compiled from properties processing input vectors $p_1,\dots,p_n$ and yielding output vectors $y_1,\dots,y_m$ at each time step.}
\label{fig:multi-monitor}
\end{figure}

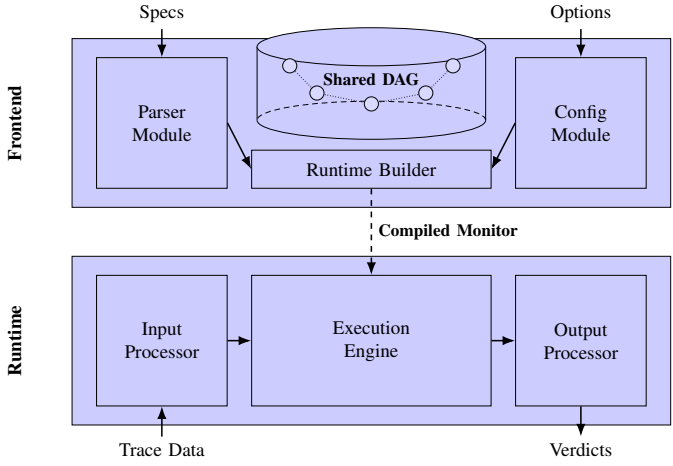
\begin{figure}[t]
\centering
\resizebox{1.0\columnwidth}{!}{%
\begin{tikzpicture}[
    >=Latex,
    outer/.style={draw, fill=blue!20, minimum width=11cm, minimum height=3.1cm},
    box/.style={draw, fill=blue!20, minimum width=2.4cm, minimum height=2.4cm, align=center},
    smallbox/.style={draw, fill=blue!20, minimum width=4.4cm, minimum height=0.7cm, align=center},
    cyl/.style={draw, fill=blue!20},
    flow/.style={->, thick}
] 

\node[outer] (top) at (0,2.2) {};

\node[rotate=90, anchor=south] at ([xshift=-0.8cm]top.west) {\textbf{Frontend}};

\node[box, anchor=west] (parser) at ([xshift=0.45cm]top.west) {Parser\\Module};

\node[box, anchor=east] (config) at ([xshift=-0.45cm]top.east) {Config\\Module};

\begin{scope}[shift={(0,3.15)}]
    \draw[cyl] (-2.1,-0.9) -- (-2.1,0.45);
    \draw[cyl] ( 2.1,-0.9) -- ( 2.1,0.45);
    \draw[cyl] (0,0.45) ellipse (2.1 and 0.35);
    \draw[cyl] (-2.1,-0.9) arc[start angle=180,end angle=360,x radius=2.1,y radius=0.35];
    \draw[densely dashed] (2.1,-0.9) arc[start angle=0,end angle=180,x radius=2.1,y radius=0.35];

    \node[font=\small\bfseries] at (0, -0.15) {Shared DAG};

    \coordinate (n1) at (-1.5, 0.1);
    \coordinate (n2) at (-1.0,-0.4);
    \coordinate (n3) at ( 1.5, 0.1);
    \coordinate (n4) at ( 1.0,-0.4);
    \coordinate (n5) at ( 0.0,-0.6);

    \draw[->, >=stealth, densely dotted] (n1) -- (n2);
    \draw[->, >=stealth, densely dotted] (n2) -- (n5);
    \draw[->, >=stealth, densely dotted] (n3) -- (n4);
    \draw[->, >=stealth, densely dotted] (n4) -- (n5);

    \filldraw[fill=blue!15] (n1) circle (0.13);
    \filldraw[fill=blue!15] (n2) circle (0.13);
    \filldraw[fill=blue!15] (n3) circle (0.13);
    \filldraw[fill=blue!15] (n4) circle (0.13);
    \filldraw[fill=blue!15] (n5) circle (0.13);
\end{scope}

\node[smallbox] (frontend) at (0,1.35) {Runtime Builder};

\draw[flow] (parser.east) -- (frontend.west);
\draw[flow] (config.west) -- (frontend.east);
\node (specs) at ([yshift=0.8cm]parser.north) {Specs};
\draw[flow] (specs) -- (parser.north);
\node (options) at ([yshift=0.8cm]config.north) {Options};
\draw[flow] (options) -- (config.north);

\node[outer] (bottom) at (0,-1.8) {};

\node[rotate=90, anchor=south] at ([xshift=-0.8cm]bottom.west) {\textbf{Runtime}};

\node[box, anchor=west] (input) at ([xshift=0.45cm]bottom.west) {Input\\Processor};

\node[draw, fill=blue!20, minimum width=4.4cm, minimum height=2.4cm, align=center] (update) at (0,-1.8) {Execution\\Engine};

\node[box, anchor=east] (output) at ([xshift=-0.45cm]bottom.east) {Output\\Processor};

\draw[flow] (input.east) -- (update.west);
\draw[flow] (update.east) -- (output.west);
\node (trace) at ([yshift=-0.8cm]input.south) {Trace Data};
\draw[flow] (trace) -- (input.south);
\node (verdicts) at ([yshift=-0.8cm]output.south) {Verdicts};
\draw[flow] (output.south) -- (verdicts);

\draw[->, thick, dashed] (frontend.south) -- (update.north) node[midway, right, font=\small\bfseries] {Compiled Monitor};

\end{tikzpicture}%
}
\caption{Architecture overview of the monitoring framework, illustrating the parsing and compilation pipeline, configuration interface, and runtime execution components.}
\label{fig:architecture-overview}
\end{figure}

This section presents the architecture and design of our multi-property monitoring framework.
Figure~\ref{fig:multi-monitor} illustrates the monitor's external interface. A collection of property specifications (shown as stacked pages on the left) is compiled into a single multi-property monitor. At each time step, the monitor receives $n$ atomic-predicate valuations $p_1, \ldots, p_n$ as input and produces $m$ property verdicts $y_1, \ldots, y_m$ as output, where $m$ is the number of registered properties.
The monitor supports both discrete-time and dense-time traces. In \emph{discrete-time} mode it is called once per timestamp and maintains scalar Boolean outputs; in \emph{dense-time} mode the monitor is presented with one time interval per call and maintains interval-valued outputs. Both modes share the same underlying infrastructure.

Figure~\ref{fig:architecture-overview} shows the internal architecture, which is divided into a \emph{frontend} and a \emph{runtime}.
As explained in Section~\ref{sec:background}, each temporal logic formula is initially translated into a tree structure mirroring its syntax. The \emph{parser module} performs this translation for all incoming property specifications. Rather than keeping these trees independent, the \emph{runtime builder} generates a unified DAG of computation nodes from the parsed formulas; this is where structural deduplication takes place, merging equivalent subformulas into shared nodes. The \emph{config module} finalizes the monitor by selecting the time model (discrete or dense).
The \emph{runtime} handles trace evaluation through a three-step pipeline. First, the \emph{input processor} handles the necessary transformations from rich message formats (including JSON records, binary formats, and raw structs) into the atomic predicate valuations expected by the monitor. Second, the \emph{execution engine} consumes these valuations and walks through the DAG nodes in topological order. As introduced in Section~\ref{sec:background}, each node is evaluated according to the update rule determined by its syntactic type, computing its new value from the current valuations of its dependencies and, for temporal operators, its previously stored state. Third, the \emph{output processor} reads the evaluated results from the root nodes of each property and delivers the final verdicts to the user.
The remainder of this section describes these components in more detail in two parts: Section~\ref{sec:frontend} details the frontend and its structural node deduplication, while Section~\ref{sec:runtime} describes the runtime along with its linearized execution schedule and double-buffered arena execution model.
\subsection{Frontend: Property Compilation}
\label{sec:frontend}

The frontend compiles a set of property specifications into a shared computation structure. Properties are specified using the Reelay Expression Format~\cite{ulus2026reelay}, a textual notation for past-time temporal logic formulas. The parser module processes each specification and constructs a syntax tree, which is then passed to the runtime builder for node generation.

\paragraph{Content-addressable node database}
Rather than building one independent computation graph per property, the runtime builder maintains a \emph{single shared DAG} across all registered properties, together with a set of designated \emph{root nodes} identifying each property's top-level formula.
Deduplication is realized through a \emph{content-addressable database} that keys each computation node by its structural signature. Each node is assigned a unique identifier, and the database maps structural hashes to these identifiers.

During bottom-up construction, before inserting any subformula node, the runtime builder computes a \emph{structural hash} that encodes the semantics of the node. The hash always includes the operator type; the remaining components depend on the node category:
for \emph{atomic predicates}, the predicate name; for \emph{unary operators} (negation), the child identifier; for \emph{unary temporal operators} (Once, Historically), the child identifier and interval bound parameters; for \emph{commutative binary operators} (conjunction, disjunction), the canonicalized pair of child identifiers, sorted so that $p \wedge q$ and $q \wedge p$ produce the same hash; and for the \emph{non-commutative binary temporal operator} (Since), the ordered pair of child identifiers and interval bound parameters.

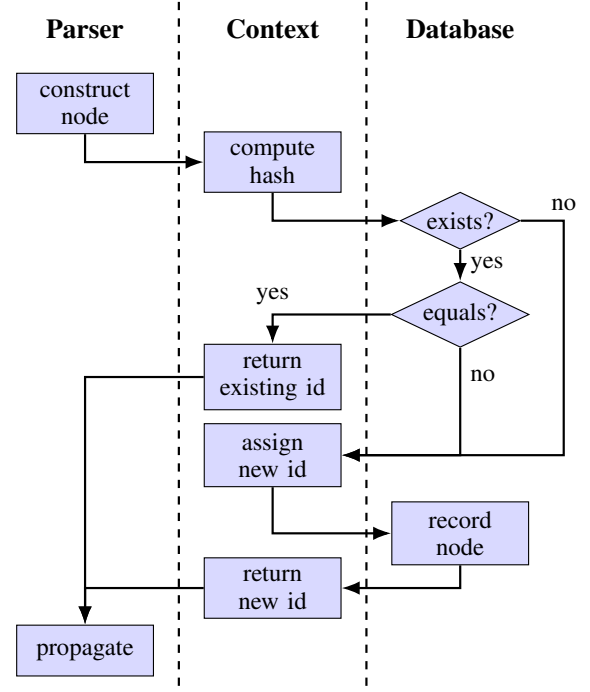
\begin{figure}[t]
\centering
\resizebox{0.85\columnwidth}{!}{%
\begin{tikzpicture}[
    >=Latex,
    font=\small,
    box/.style={draw, fill=blue!15, minimum width=1.75cm, minimum height=0.65cm, align=center},
    decision/.style={draw, diamond, fill=blue!15, aspect=2.1, inner sep=1pt, align=center},
    arr/.style={->, thick},
    lane/.style={dashed, thick}
]

\def\ytop{0.8}
\def\ybot{-8}
\def\xA{-1.7}
\def\xB{0.7}
\def\xC{3.1}
\def\xD{5.5}

\draw[lane] (\xB,\ytop) -- (\xB,\ybot);
\draw[lane] (\xC,\ytop) -- (\xC,\ybot);

\node[font=\bfseries] at (-0.5,0.45) {Parser};
\node[font=\bfseries] at (1.9,0.45) {Context};
\node[font=\bfseries] at (4.3,0.45) {Database};

\def\xP{-0.5}
\def\xK{1.9}
\def\xDbase{4.3}

\node[box]      (create) at (\xP,-0.5) {construct\\node};
\node[box]      (hash)   at (\xK,-1.25) {compute\\hash};
\node[decision] (exists) at (\xDbase,-2.0) {exists?};
\node[decision] (equals) at (\xDbase,-3.2) {equals?};

\node[box]      (oldid)  at (\xK,-4.0) {return\\existing id};
\node[box]      (assign) at (\xK,-5.0) {assign\\new id};
\node[box]      (newid)  at (\xK,-6.7) {return\\new id};

\node[box]      (record) at (\xDbase,-6) {record\\node};
\node[box]      (prop)   at (\xP,-7.5) {propagate};

\draw[arr] (create.south) |- (hash.west);
\draw[arr] (hash.south) |- (exists.west);

\draw[arr] (exists) -- node[right] {yes} (equals);

\draw[arr] (equals) -| node[above] {yes} (oldid);
\draw[arr] (oldid.west) -| (prop.north);

\draw[arr] (equals.south) -- ++(0,-0.35) node[right] {no} |- (assign.east);
\draw[arr] (exists.east) -- ++(0.55,0) node[above] {no} |- (assign.east);

\draw[arr] (assign) |- (record);
\draw[arr] (record.south) |- (newid.east);
\draw[arr] (newid.west) -| (prop.north);

\end{tikzpicture}%
}
\caption{Node creation and lookup workflow across parser, context, and content-addressable database components.}
\label{fig:resource-workflow}
\end{figure}

\paragraph{Integration into the parsing process}
Figure~\ref{fig:resource-workflow} shows the node creation and lookup workflow. During parsing, each subformula is constructed bottom-up. The parser creates a node description and passes it to a context layer, which computes the structural hash. The hash is looked up in the content-addressable database: if a matching entry exists and structural equality is confirmed, the existing identifier is returned and reused; otherwise, a new identifier is assigned and the node is recorded in the database. The returned identifier is then propagated upward to serve as a child reference in parent nodes.

\begin{figure}[t]
\centering
\resizebox{0.55\columnwidth}{!}{%
\begin{tikzpicture}[
    >=Latex,
    every node/.style={font=\small},
    dagnode/.style={draw, circle, thick, minimum size=8mm, inner sep=0pt},
    shared/.style={fill=green!25},
    root1/.style={fill=red!25},
    root2/.style={fill=blue!25},
    edge/.style={->, thick}
]

\node[dagnode, shared] (p) at (0, 0) {$p$};
\node[dagnode, shared] (q) at (2.4, 0) {$q$};

\node[dagnode, shared] (andpq) at (1.2, 1.0) {$\wedge$};

\node[dagnode, root1] (once) at (-0.2, 2.2) {$\once_{[0,10]}$};
\node[dagnode, root2] (hist) at (2.6, 2.2) {$H$};

\draw[edge] (p) -- (andpq);
\draw[edge] (q) -- (andpq);
\draw[edge] (andpq) -- (once);
\draw[edge] (andpq) -- (hist);

\node[font=\footnotesize, above=2pt] at (once.north) {$\mathit{root}_1$ ($\varphi_1$)};
\node[font=\footnotesize, above=2pt] at (hist.north) {$\mathit{root}_2$ ($\varphi_2$)};

\end{tikzpicture}%
}
\caption{Structural deduplication: two properties $\varphi_1 = \once_{[0,10]}(p \wedge q)$ and $\varphi_2 = H(p \wedge q)$ compiled into a shared DAG. Green nodes are shared; colored roots are property-specific.}
\label{fig:dedup-example}
\end{figure}
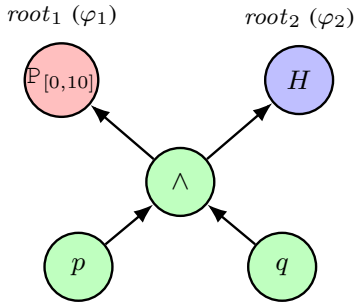

Consider two properties $\varphi_1 = \once_{[0,10]}({p \wedge q})$ and $\varphi_2 = H({p \wedge q})$. Figure~\ref{fig:dedup-example} illustrates how the two formula trees merge into a single shared DAG. The atomic predicates $p$ and $q$ and the conjunction $p \wedge q$ are shared between both properties; only the two distinct root operators are unique. The resulting shared DAG contains five nodes rather than the eight that would result from independent compilation.

The sub-expression $(p \wedge q)$ is evaluated only once per timestep, regardless of how many properties reference it. A single pass through the shared DAG in topological order computes all property outputs simultaneously; results are read directly from each property's root node.

\paragraph{Configuration and linearized execution schedule}
Properties may be registered incrementally; each registration appends newly created nodes to the shared DAG and records the new root index. After all properties are registered, the config module finalizes the monitor by selecting the time model. At this stage, the compiler serializes the shared DAG into a \emph{linearized execution schedule}: a contiguous array of node records arranged in strict topological order (operands before operators). Instead of generating heap-allocated objects connected by memory pointers, the entire formula graph is packed into a single contiguous memory region where child references are simple integer array indices. This ordering enables the runtime to evaluate the entire graph in a single forward pass without recursion. Once finalized, the execution schedule is immutable and the monitor begins accepting trace data.

\subsection{Runtime: Data-Oriented Execution Model}
\label{sec:runtime}

The runtime adopts a \emph{data-oriented design} that eliminates heap allocations for intermediate partial results during the monitoring loop. As discussed in Section~\ref{sec:background}, classical object-oriented implementations rely on heap-allocated nodes connected by pointers, leading to fragmented memory and irregular access patterns during tree traversal. Because our frontend instead pre-compiles the DAG into a linearized execution schedule (as described in Section~\ref{sec:frontend}), the runtime is structured to benefit from spatial locality.

However, beyond the nodes themselves, storing intermediate interval sets as heap-allocated objects adds significant allocation pressure inside the hot evaluation loop. To resolve this, the framework relies on a zero-allocation double-buffered arena.
\subsubsection*{Zero-Allocation Double-Buffered Arena}
\label{sec:arena}

\begin{figure}[t]
\centering

\begin{subfigure}[t]{\columnwidth}
\centering
\begin{tikzpicture}

\def\cell{0.25}
\def\cols{20}

\foreach \start/\stop/\clr in {
    0/9/red!30,
    10/23/blue!30,
    24/48/orange!30,
    49/59/green!30,
    60/78/purple!30
}{
    \foreach \i in {\start,...,\stop} {
        \pgfmathtruncatemacro{\col}{mod(\i,\cols)}
        \pgfmathtruncatemacro{\row}{int(\i/\cols)}
        \fill[\clr] (\col*\cell, 2.5-\row*\cell) rectangle ++(\cell,\cell);
    }
}

\foreach \start/\stop/\clr in {
    0/8/red!30,
    9/25/blue!30,
    26/43/orange!30
}{
    \foreach \i in {\start,...,\stop} {
        \pgfmathtruncatemacro{\col}{mod(\i,\cols)}
        \pgfmathtruncatemacro{\row}{int(\i/\cols)}
        \fill[\clr] (\col*\cell, 1.0-\row*\cell) rectangle ++(\cell,\cell);
    }
}

\draw[step=\cell, thick, black] (0,1.5) grid (5,2.75);
\draw[thick, black] (0,1.5) rectangle (5,2.75);
\draw[step=\cell, thick, black] (0,0) grid (5,1.25);
\draw[thick, black] (0,0) rectangle (5,1.25);

\node[anchor=west] at (5.1,2.1) {\small $B_{\mathrm{prev}}$};
\node[anchor=west] at (5.1,0.6) {\small $B_{\mathrm{cur}}$};

\pgfmathtruncatemacro{\wIdx}{44}
\pgfmathtruncatemacro{\wCol}{mod(\wIdx,\cols)}
\pgfmathtruncatemacro{\wRow}{int(\wIdx/\cols)}
\coordinate (wcell) at (\wCol*\cell + 0.5*\cell, 1.0 - \wRow*\cell + 0.5*\cell);
\node[anchor=north] (wlabel) at (\wCol*\cell + 0.5*\cell, -0.1) {\scriptsize $\mathit{idx}_{\mathrm{write}}$};
\draw[-Latex, very thick, green!60!black] (wlabel.north) -- (wcell);

\node[anchor=east] (rnode) at (-0.1,2.1) {};
\node[anchor=east] (wnode) at (-0.1,0.6) {};
\draw[<->, very thick] (rnode) to[out=160, in=200] (wnode);
\node[rotate=90, anchor=south] at (-0.9,1.35) {\scriptsize swap};

\end{tikzpicture}
\caption{Global buffer roles and swap mechanism. The previous buffer ($B_{\mathrm{prev}}$) holds the previous timestep's states; the current buffer ($B_{\mathrm{cur}}$) accumulates the current timestep's states. Buffers are swapped at each timestep boundary.}
\label{fig:arena-diagram}
\end{subfigure}

\vspace{0.2cm}

\begin{subfigure}[t]{\columnwidth}
\centering
\begin{tikzpicture}

\def\cell{0.25}
\def\cols{20}
\def\rows{5}

\foreach \start/\stop/\clr in {
    0/9/red!30,
    10/23/blue!30,
    24/48/gray!30
}{
    \foreach \i in {\start,...,\stop} {
        \pgfmathtruncatemacro{\col}{mod(\i,\cols)}
        \pgfmathtruncatemacro{\row}{int(\i/\cols)}
        \fill[\clr] (\col*\cell, 0-\row*\cell) rectangle ++(\cell,\cell);
    }
}

\draw[step=\cell, thick, black] (0,-0.5) grid (5,0.25);

\pgfmathtruncatemacro{\rOne}{0}   
\pgfmathtruncatemacro{\rTwo}{10}   
\pgfmathtruncatemacro{\wIdx}{49}   

\pgfmathtruncatemacro{\rOneCol}{mod(\rOne,\cols)}
\pgfmathtruncatemacro{\rOneRow}{int(\rOne/\cols)}

\pgfmathtruncatemacro{\rTwoCol}{mod(\rTwo,\cols)}
\pgfmathtruncatemacro{\rTwoRow}{int(\rTwo/\cols)}

\pgfmathtruncatemacro{\wCol}{mod(\wIdx,\cols)}
\pgfmathtruncatemacro{\wRow}{int(\wIdx/\cols)}

\coordinate (r1cell) at (\rOneCol*\cell + 0.5*\cell, -\rOneRow*\cell + 0.5*\cell);
\coordinate (r2cell) at (\rTwoCol*\cell + 0.5*\cell, -\rTwoRow*\cell + 0.5*\cell);
\coordinate (wcell)  at (\wCol*\cell    + 0.5*\cell, -\wRow*\cell    + 0.5*\cell);

\node[anchor=south] (r1) at ($(r1cell)+(0,0.25)$) {\scriptsize read};
\node[anchor=south] (r2) at ($(r2cell)+(0,0.25)$) {\scriptsize read};
\node[anchor=north] (w)  at ($(wcell)+(0,-0.25)$) {\scriptsize write};

\draw[-Latex, very thick, red]
    (r1.south) -- (r1cell);

\draw[-Latex, very thick, blue]
    (r2.south) -- (r2cell);

\draw[-Latex, very thick, green!60!black]
    (w.north) -- (wcell);

\end{tikzpicture}
\caption{Initial position of pointers before the current node evaluation}
\label{fig:buffer-read-write-a}
\end{subfigure}

\begin{subfigure}[t]{\columnwidth}
\centering
\begin{tikzpicture}

\def\cell{0.25}
\def\cols{20}
\def\rows{5}

\foreach \start/\stop/\clr in {
    0/9/red!30,
    10/23/blue!30,
    24/48/gray!30,
    49/54/green!30
}{
    \foreach \i in {\start,...,\stop} {
        \pgfmathtruncatemacro{\col}{mod(\i,\cols)}
        \pgfmathtruncatemacro{\row}{int(\i/\cols)}
        \fill[\clr] (\col*\cell, 0-\row*\cell) rectangle ++(\cell,\cell);
    }
}

\draw[step=\cell, thick, black] (0,-0.5) grid (5,0.25);

\pgfmathtruncatemacro{\rOne}{7}   
\pgfmathtruncatemacro{\rTwo}{17}   
\pgfmathtruncatemacro{\wIdx}{55}   

\pgfmathtruncatemacro{\rOneCol}{mod(\rOne,\cols)}
\pgfmathtruncatemacro{\rOneRow}{int(\rOne/\cols)}

\pgfmathtruncatemacro{\rTwoCol}{mod(\rTwo,\cols)}
\pgfmathtruncatemacro{\rTwoRow}{int(\rTwo/\cols)}

\pgfmathtruncatemacro{\wCol}{mod(\wIdx,\cols)}
\pgfmathtruncatemacro{\wRow}{int(\wIdx/\cols)}

\coordinate (r1cell) at (\rOneCol*\cell + 0.5*\cell, -\rOneRow*\cell + 0.5*\cell);
\coordinate (r2cell) at (\rTwoCol*\cell + 0.5*\cell, -\rTwoRow*\cell + 0.5*\cell);
\coordinate (wcell)  at (\wCol*\cell    + 0.5*\cell, -\wRow*\cell    + 0.5*\cell);

\node[anchor=south] (r1) at ($(r1cell)+(0,0.25)$) {\scriptsize read};
\node[anchor=south] (r2) at ($(r2cell)+(0,0.25)$) {\scriptsize read};
\node[anchor=north] (w)  at ($(wcell)+(0,-0.25)$) {\scriptsize write};

\draw[-Latex, very thick, red]
    (r1.south) -- (r1cell);

\draw[-Latex, very thick, blue]
    (r2.south) -- (r2cell);

\draw[-Latex, very thick, green!60!black]
    (w.north) -- (wcell);

\end{tikzpicture}
\caption{Progression of pointers during the current node evaluation}
\label{fig:buffer-read-write-b}
\end{subfigure}
\caption{Zero-allocation double-buffered arena. (a) Global buffer roles and swap mechanism. (b) Initial position of pointers. (c) Progression of pointers.}
\label{fig:arena-combined}
\end{figure}
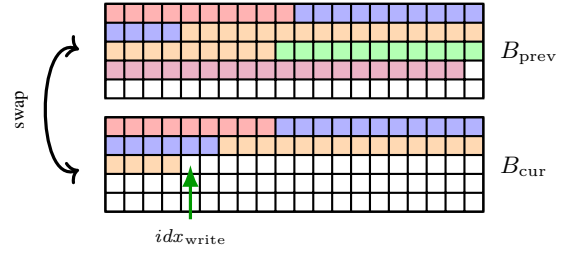
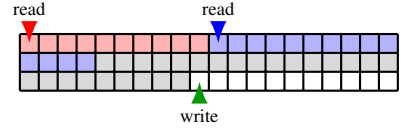
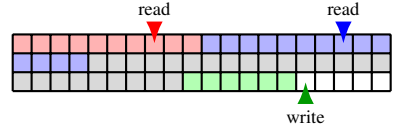

Each monitoring operator produces or consumes an \emph{interval set} representing the set of time points for which a sub-formula could hold. Rather than scattering these sets across individual heap-allocated objects, the framework manages all interval-set data using a \emph{double-buffered arena} consisting of two contiguous memory buffers. 

As illustrated in Figure~\ref{fig:arena-combined}a, the arena maintains a \emph{previous buffer} ($B_{\mathrm{prev}}$) holding the accumulated state from the previous timestep, and a \emph{current buffer} ($B_{\mathrm{cur}}$) into which the current timestep's state is accumulated. At the end of each timestep, the roles of the two buffers are swapped and the write cursor is reset to zero. As a result, the newly written states become the previous-timestep history, and the new current buffer is effectively cleared for the next evaluation.

While Figure~\ref{fig:arena-combined}a shows the global state of both buffers, Figure~\ref{fig:arena-combined}(b-c) zooms in exclusively on $B_{\mathrm{cur}}$ to illustrate the internal access pattern during a single evaluation step. Because nodes are evaluated in topological order, a node's children have already been evaluated and their states occupy earlier regions of this current buffer (shown in color). The node reads these regions via read pointers, computes its result, and appends its updated state at the write cursor; the advancing cursor then allocates a fresh interval-set handle for the evaluated node. Temporal operators such as Since additionally read their own accumulated state from $B_{\mathrm{prev}}$.

The consequence is that \emph{no dynamic memory allocation takes place during any evaluation timestep}: all interval data occupies a single contiguous memory block, and the two arena buffers are provisioned exactly once at monitor creation based on a strict static upper bound, detailed in Section~\ref{sec:implementation}.

\section{Implementation and Evaluation}
\label{sec:evaluation}
This section outlines implementation details of our multi-property monitoring framework, \emph{LoomRV}, and presents our experimental results.
Throughout this section, we compare \emph{LoomRV} against \emph{Reelay}~\cite{ulus2026reelay}, a state-of-the-art online runtime monitor for past-time LTL and MTL. 
Reelay follows the same compositional sequential-network construction~\cite{ulus2026online} as the single-property baseline of Section~\ref{sec:background}, but synthesizes one independent monitor per property and uses an object-oriented design.
We first detail how the linearized execution schedule is concretely laid out and how the arena's memory capacity is statically bounded, and describe the common hardware and measurement setup used in every subsequent experiment.
Sections~\ref{sec:eval-cse},~\ref{sec:eval-single}, and~\ref{sec:eval-multi} then evaluate the impact of cross-property sharing on the shared-DAG speedup, the per-property evaluation cost against Reelay, and multi-property throughput, respectively.
\label{sec:implementation}

\emph{LoomRV} is implemented in C++. Its modules mirror the architecture of Figure~\ref{fig:architecture-overview}: the \emph{Parser} and \emph{Runtime Builder} jointly compile \PastMTL\ specifications into the linearized execution schedule of Section~\ref{sec:frontend}, the \emph{Execution Engine} evaluates that array against incoming traces using the interval set implementation with the double-buffered arena of Section~\ref{sec:runtime}, and the \emph{Input Processor} ingests trace data in one of two interchangeable realizations: a \emph{JSON mode} that parses newline-delimited JSON records with a SIMD-accelerated JSON parser~\cite{langdale2019parsing}, and a \emph{binary mode} that reads a compact pre-encoded bitfield format, eliminating JSON parsing overhead from the hot path.

\textbf{Execution Schedule Layout.}
\label{sec:impl-node-array}
Concretely, each entry of the linearized execution schedule is a fixed-size record holding
(i)~the operator tag (proposition, negation, conjunction, disjunction, Previous, Since, Once, Historically, \ldots),
(ii)~two integer child indices into the same array,
(iii)~the operator's temporal bound parameters,
(iv)~an interval-set-valued \emph{persistent state} carried across timesteps, and
(v)~an interval-set-valued \emph{current output} written in the present timestep.
In discrete-time mode the current output degrades to a scalar Boolean while the persistent state remains an interval set; in dense-time mode both fields are interval-set handles into the double-buffered arena.

\textbf{Buffer Sizing and Memory Limits.}
Because the arena buffers are fully reset at each timestep boundary, the required memory capacity is strictly determined by the maximum fragmentation of the interval sets evaluated during a single pass. For predicate, Boolean, and untimed temporal nodes, the output is a single contiguous interval or scalar state, requiring $\mathcal{O}(1)$ space. For metric temporal nodes, the memory footprint is dictated by the future temporal marking mechanism~\cite{ulus2026online}, which maintains a set of future intervals within a sliding window bounded by the operator's upper time constraint $b$.

The worst-case memory limit is reached when this valuation set undergoes maximum fragmentation. This occurs under adversarial conditions, for example, when a highly alternating input signal is evaluated against a point-interval property such as $\once_{[b:b]} q$. In this scenario, marked time points are completely disjoint and non-adjacent. Within a window of size $b$, the maximum number of strictly non-overlapping, non-adjacent points is exactly $\lceil b/2 \rceil$. Consequently, a timed node must store at most $\mathcal{O}(b)$ distinct intervals. For a shared DAG containing $N$ nodes with a maximum upper time bound $B$, the total required buffer capacity is strictly bounded by $\mathcal{O}(N \cdot B)$. Because this worst-case space complexity depends entirely on the static temporal parameters of the compiled formulas rather than the trace length, the framework computes this upper bound and allocates the arena exactly once at monitor creation, guaranteeing zero dynamic allocations and zero overflow overhead during runtime evaluation. It is worth noting that in our experiments, actual buffer consumption remained well below this theoretical maximum. In typical monitoring scenarios, overlapping temporal markings naturally merge into a small number of contiguous blocks, ensuring the memory footprint remains highly compact in practice.

\textbf{Experimental Setup.}
\label{sec:setup}
In all our experiments, we report performance results from our Linux-based containerized benchmarking environment, which was run on an Intel Core i7-10750H (6 cores, 2.60\,GHz base, 12\,MB L3 cache).

The trace corpora and comparison baselines differ across experiments and are described within each subsection below.

\subsection{Impact of Cross-Property Sharing}
\label{sec:eval-cse}
To quantify how much the speedup of multi-property monitoring depends on the degree of structural sharing among formulas, we designed four synthetic formula sets with known sharing characteristics.
We measure this sharing in terms of node compression: the ratio of total nodes required for independent monitors to the number of nodes in the unified shared DAG.
Each set contains 10 \PastMTL\ formulas evaluated over the same 1\,000\,000-step traces in both discrete and dense time.
We compare three configurations: Reelay Sequential (one Reelay monitor per formula), LoomRV Sequential (one independent LoomRV monitor per formula), and LoomRV Multi (a single unified monitor).

We evaluate the following four scenarios:
\begin{itemize}
  \item \textbf{Best-case shared core.}
        All 10 formulas take the form $(\varphi_{\mathit{core}} \vee \once_{[:k]}(p))$ where $\varphi_{\mathit{core}} = H(\once_{[:10]}(q) \to (\neg p \mathcal{S} r))$ is identical across all properties, and only the time bound $k \in \{1,\ldots,10\}$ of the small $\once$ term differs.
        Only the 10-node $\varphi_{\mathit{core}}$ subtree is shared across all 10 properties.
        This yields a \textbf{3.57$\times$} node compression (72\% of nodes deduplicated).
  \item \textbf{Nested best-case.}
        Property $k$ is $H(L_k)$ where $L_k = (L_{k-1} \vee s_k)$, $s_k \in \{p, q, r\}$ cycling with $k$, and $L_1 = (p \vee q)$.
        Since $L_{k-1}$ appears identically as a subtree in formula $k$, structural deduplication reuses every previously compiled $\vee$ subexpression as a prefix, yielding \textbf{4.09$\times$} node compression (75\% of nodes deduplicated).
  \item \textbf{Worst-case unique leaves.}
        Formulas share a structural template but carry distinct temporal bounds on every operator, preventing bound-sensitive deduplication.
        Only propositions and a few structural operators are shared, resulting in \textbf{1.67$\times$} node compression (40\% of nodes deduplicated).
  \item \textbf{Nested worst-case.}
        Escalating AND chains with unique \texttt{once} bounds at every nesting level, so almost every node is distinct.
        This offers minimal sharing, yielding just \textbf{1.31$\times$} node compression (23\% of nodes deduplicated).
\end{itemize}

\begin{table}[htbp]
    \centering
    \caption{Impact of cross-property sharing (discrete): execution time (s) for four scenarios with 10 properties over 1\,000\,000 timesteps.}
    \label{tab:cse-sensitivity-discrete}
    \resizebox{\columnwidth}{!}{
    \begin{tabular}{l r r r r}
        \toprule
        \textbf{Scenario} & \textbf{Comp.} & \textbf{Reelay Seq} & \textbf{LoomRV Seq} & \textbf{LoomRV Multi} \\
        \midrule
        \multicolumn{5}{l}{\textit{JSON feeder}} \\
        \midrule
        Best-case shared   & 3.57$\times$ & 5.09 & 1.39 & \textbf{0.252} \\
        Nested best-case   & 4.09$\times$ & 3.32 & 1.10 & \textbf{0.206} \\
        Worst-case unique  & 1.67$\times$ & 3.62 & 1.33 & \textbf{0.420} \\
        Nested worst-case  & 1.31$\times$ & 4.40 & 1.23 & \textbf{0.385} \\
        \midrule
        \multicolumn{5}{l}{\textit{Binary feeder}} \\
        \midrule
        Best-case shared   & 3.57$\times$ & 2.80 & 0.61 & \textbf{0.175} \\
        Nested best-case   & 4.09$\times$ & 1.34 & 0.39 & \textbf{0.127} \\
        Worst-case unique  & 1.67$\times$ & 1.98 & 0.52 & \textbf{0.346} \\
        Nested worst-case  & 1.31$\times$ & 2.23 & 0.50 & \textbf{0.340} \\
        \bottomrule
    \end{tabular}
    }
\end{table}

\begin{table}[htbp]
    \centering
    \caption{Impact of cross-property sharing (dense): execution time (s) for four scenarios with 10 properties over 1\,000\,000 timesteps.}
    \label{tab:cse-sensitivity}
    \resizebox{\columnwidth}{!}{
    \begin{tabular}{l r r r r}
        \toprule
        \textbf{Scenario} & \textbf{Comp.} & \textbf{Reelay Seq} & \textbf{LoomRV Seq} & \textbf{LoomRV Multi} \\
        \midrule
        \multicolumn{5}{l}{\textit{JSON feeder}} \\
        \midrule
        Best-case shared   & 3.57$\times$ & 18.08 & 3.61 & \textbf{0.850} \\
        Nested best-case   & 4.09$\times$ & 7.44 & 2.16 & \textbf{0.604} \\
        Worst-case unique  & 1.67$\times$ & 9.16 & 3.03 & \textbf{1.792} \\
        Nested worst-case  & 1.31$\times$ & 8.69 & 2.69 & \textbf{1.476} \\
        \midrule
        \multicolumn{5}{l}{\textit{Binary feeder}} \\
        \midrule
        Best-case shared   & 3.57$\times$ & 12.26 & 2.47 & \textbf{0.715} \\
        Nested best-case   & 4.09$\times$ & 5.63 & 1.12 & \textbf{0.490} \\
        Worst-case unique  & 1.67$\times$ & 7.52 & 1.97 & \textbf{1.576} \\
        Nested worst-case  & 1.31$\times$ & 6.84 & 1.55 & \textbf{1.286} \\
        \bottomrule
    \end{tabular}
    }
\end{table}

Tables~\ref{tab:cse-sensitivity-discrete} and~\ref{tab:cse-sensitivity} report the results. The data confirm that multi-property speedup scales directly with the degree of structural sharing across both trace models.

In \textbf{discrete-time} (Table~\ref{tab:cse-sensitivity-discrete}), at maximum sharing (nested best-case, $4.09\times$ compression), LoomRV Multi is $16.1\times$ (JSON) and $10.6\times$ (binary) faster than Reelay Sequential, and $5.3\times$ (JSON) and $3.1\times$ (binary) faster than LoomRV Sequential. At minimal sharing (nested worst-case, $1.31\times$ compression), Multi remains $11.4\times$ (JSON) and $6.6\times$ (binary) faster than Reelay Sequential, and $3.2\times$ (JSON) and $1.5\times$ (binary) faster than LoomRV Sequential. 
Notably, achieving up to a $3.2\times$ speedup over LoomRV Sequential despite only a $1.31\times$ reduction in nodes highlights a critical characteristic of discrete-time evaluation: because individual node processing is computationally trivial (scalar Booleans), execution time is dominated by program initialization and trace parsing overhead. Since the sequential baseline must initialize the framework and parse the trace 10 separate times, the multi-property monitor achieves a massive speedup simply by ingesting the trace once and distributing the inputs to all properties simultaneously. This also explains why the relative speedup multipliers are lower for the binary feeder: because binary parsing is highly optimized, the sequential baseline wastes far less time on I/O. Consequently, avoiding redundant parsing passes yields a smaller relative multiplier, even though the binary monitor achieves the fastest absolute execution times overall.

In \textbf{dense-time} (Table~\ref{tab:cse-sensitivity}), the same trends hold: at maximum sharing, LoomRV Multi is $12.3\times$ (JSON) and $11.5\times$ (binary) faster than Reelay Sequential, and $3.6\times$ (JSON) and $2.3\times$ (binary) faster than LoomRV Sequential. At minimal sharing, the advantage over LoomRV Sequential shrinks to $1.8\times$ (JSON) and $1.2\times$ (binary). Because dense-time node processing involves heavier interval-set operations rather than simple Booleans, the evaluation loop demands a larger share of the overall execution time. Consequently, parsing overhead becomes less dominant, and the multi-property speedups align more closely with the actual structural compression ratio. Yet, Multi still remains $5.9\times$ (JSON) and $5.3\times$ (binary) faster than Reelay Sequential, as the linearized-schedule evaluator and zero-allocation arena continue to provide substantial baseline efficiency.

\subsection{Single-Property Comparison}
\label{sec:eval-single}
To isolate the per-property evaluation cost from multi-property effects, we compare LoomRV against Reelay on each of the 30 properties independently.
Here, each property is evaluated on its own dedicated trace generated by the \textit{timescales} benchmark generator \cite{ulus2019timescales}. The tool synthesizes traces with tunable event frequencies, allowing us to evaluate the monitors under varying degrees of temporal density. Each generated trace spans a maximum simulated time of 1,000,000.
We verified that none of the 30 \textit{timescales} formulas contain repeated subexpressions, so no intra-formula deduplication occurs during compilation.

Table~\ref{tab:discrete_results} reports discrete-time results using both the JSON and binary feeders. In discrete-time mode, LoomRV maintains a consistent advantage with a median speedup of $2.00\times$ (JSON feeder, mean $2.01\times$) and $1.81\times$ (binary feeder, mean $2.01\times$), with individual speedups ranging from $1.55\times$ to $2.82\times$.
Tables~\ref{tab:dense_json} and~\ref{tab:dense_binary} report the corresponding dense-time results across three temporal densities (Dense1, Dense10, Dense100).
Across all 90 dense-time configurations, LoomRV consistently outperforms Reelay: the median per-configuration speedup is $3.45\times$ (JSON feeder, mean $3.46\times$) and $4.53\times$ (binary feeder, mean $4.43\times$), with individual speedups ranging from $2.2\times$ to $5.8\times$ depending on the formula and timescale granularity.
Across both trace models, the binary feeder yields additional speedup over the JSON feeder for both tools, as it eliminates JSON parsing overhead from the hot path.

\begin{table}[htbp]
\centering
\footnotesize
\caption{\small Execution time (s) for single-property discrete-time monitoring. Reelay vs.\ LoomRV using JSON and binary feeders.}
\label{tab:discrete_results}
\begin{tabular}{l rr rr}
\toprule
& \multicolumn{2}{c}{\textbf{JSON Feeder (s)}} & \multicolumn{2}{c}{\textbf{Binary Feeder (s)}} \\
\cmidrule(r){2-3} \cmidrule(l){4-5}
\textbf{Benchmark} & \textbf{Reelay} & \textbf{LoomRV} & \textbf{Reelay} & \textbf{LoomRV} \\
\midrule
AbsentAQ10    & 0.154 & \textbf{0.088} & 0.070 & \textbf{0.039} \\
AbsentAQ100   & 0.143 & \textbf{0.086} & 0.061 & \textbf{0.038} \\
AbsentAQ1000  & 0.142 & \textbf{0.087} & 0.060 & \textbf{0.038} \\
AbsentBQR10   & 0.243 & \textbf{0.118} & 0.088 & \textbf{0.050} \\
AbsentBQR100  & 0.217 & \textbf{0.114} & 0.071 & \textbf{0.050} \\
AbsentBQR1000 & 0.216 & \textbf{0.114} & 0.067 & \textbf{0.050} \\
AbsentBR10    & 0.163 & \textbf{0.082} & 0.090 & \textbf{0.032} \\
AbsentBR100   & 0.165 & \textbf{0.082} & 0.089 & \textbf{0.032} \\
AbsentBR1000  & 0.162 & \textbf{0.082} & 0.089 & \textbf{0.032} \\
AlwaysAQ10    & 0.154 & \textbf{0.088} & 0.068 & \textbf{0.037} \\
AlwaysAQ100   & 0.145 & \textbf{0.086} & 0.060 & \textbf{0.037} \\
AlwaysAQ1000  & 0.142 & \textbf{0.087} & 0.059 & \textbf{0.037} \\
AlwaysBQR10   & 0.236 & \textbf{0.111} & 0.082 & \textbf{0.043} \\
AlwaysBQR100  & 0.219 & \textbf{0.105} & 0.069 & \textbf{0.042} \\
AlwaysBQR1000 & 0.211 & \textbf{0.105} & 0.066 & \textbf{0.042} \\
AlwaysBR10    & 0.164 & \textbf{0.081} & 0.088 & \textbf{0.031} \\
AlwaysBR100   & 0.164 & \textbf{0.082} & 0.088 & \textbf{0.032} \\
AlwaysBR1000  & 0.162 & \textbf{0.081} & 0.088 & \textbf{0.032} \\
RecurBQR10    & 0.281 & \textbf{0.117} & 0.109 & \textbf{0.052} \\
RecurBQR100   & 0.255 & \textbf{0.114} & 0.089 & \textbf{0.051} \\
RecurBQR1000  & 0.252 & \textbf{0.114} & 0.086 & \textbf{0.051} \\
RecurGLB10    & 0.126 & \textbf{0.064} & 0.076 & \textbf{0.028} \\
RecurGLB100   & 0.098 & \textbf{0.061} & 0.051 & \textbf{0.028} \\
RecurGLB1000  & 0.094 & \textbf{0.061} & 0.048 & \textbf{0.028} \\
RespondBQR10  & 0.369 & \textbf{0.147} & 0.131 & \textbf{0.062} \\
RespondBQR100 & 0.342 & \textbf{0.145} & 0.102 & \textbf{0.063} \\
RespondBQR1000& 0.333 & \textbf{0.142} & 0.100 & \textbf{0.064} \\
RespondGLB10  & 0.242 & \textbf{0.093} & 0.115 & \textbf{0.042} \\
RespondGLB100 & 0.205 & \textbf{0.091} & 0.083 & \textbf{0.041} \\
RespondGLB1000& 0.201 & \textbf{0.090} & 0.078 & \textbf{0.041} \\
\bottomrule
\end{tabular}
\end{table}

\begin{table}[htbp]
\centering
\footnotesize
\caption{\small Execution time (s): Reelay JSON vs.\ LoomRV JSON, single-property, dense-time.}
\label{tab:dense_json}
\resizebox{\columnwidth}{!}{
\begin{tabular}{l rr rr rr}
\toprule
& \multicolumn{2}{c}{\textbf{Dense1 (s)}} & \multicolumn{2}{c}{\textbf{Dense10 (s)}} & \multicolumn{2}{c}{\textbf{Dense100 (s)}} \\
\cmidrule(r){2-3} \cmidrule(r){4-5} \cmidrule(r){6-7}
\textbf{Benchmark} & \textbf{Reelay} & \textbf{LoomRV} & \textbf{Reelay} & \textbf{LoomRV} & \textbf{Reelay} & \textbf{LoomRV} \\
\midrule
AbsentAQ10    & 0.447 & \textbf{0.131} & 0.260 & \textbf{0.075} & 0.248 & \textbf{0.073} \\
AbsentAQ100   & 0.401 & \textbf{0.122} & 0.204 & \textbf{0.063} & 0.172 & \textbf{0.053} \\
AbsentAQ1000  & 0.395 & \textbf{0.121} & 0.194 & \textbf{0.061} & 0.164 & \textbf{0.052} \\
AbsentBQR10   & 0.495 & \textbf{0.202} & 0.358 & \textbf{0.135} & 0.349 & \textbf{0.137} \\
AbsentBQR100  & 0.339 & \textbf{0.148} & 0.109 & \textbf{0.044} & 0.065 & \textbf{0.022} \\
AbsentBQR1000 & 0.316 & \textbf{0.142} & 0.078 & \textbf{0.031} & 0.029 & \textbf{0.007} \\
AbsentBR10    & 0.361 & \textbf{0.108} & 0.219 & \textbf{0.063} & 0.215 & \textbf{0.062} \\
AbsentBR100   & 0.334 & \textbf{0.104} & 0.190 & \textbf{0.056} & 0.171 & \textbf{0.049} \\
AbsentBR1000  & 0.337 & \textbf{0.103} & 0.187 & \textbf{0.055} & 0.169 & \textbf{0.049} \\
AlwaysAQ10    & 0.462 & \textbf{0.127} & 0.254 & \textbf{0.073} & 0.246 & \textbf{0.070} \\
AlwaysAQ100   & 0.406 & \textbf{0.118} & 0.195 & \textbf{0.061} & 0.158 & \textbf{0.051} \\
AlwaysAQ1000  & 0.396 & \textbf{0.117} & 0.187 & \textbf{0.059} & 0.154 & \textbf{0.050} \\
AlwaysBQR10   & 0.694 & \textbf{0.186} & 0.482 & \textbf{0.123} & 0.482 & \textbf{0.123} \\
AlwaysBQR100  & 0.471 & \textbf{0.138} & 0.149 & \textbf{0.041} & 0.082 & \textbf{0.020} \\
AlwaysBQR1000 & 0.449 & \textbf{0.128} & 0.101 & \textbf{0.029} & 0.034 & \textbf{0.007} \\
AlwaysBR10    & 0.374 & \textbf{0.105} & 0.219 & \textbf{0.062} & 0.212 & \textbf{0.060} \\
AlwaysBR100   & 0.349 & \textbf{0.100} & 0.185 & \textbf{0.056} & 0.164 & \textbf{0.048} \\
AlwaysBR1000  & 0.345 & \textbf{0.099} & 0.183 & \textbf{0.054} & 0.157 & \textbf{0.047} \\
RecurBQR10    & 0.647 & \textbf{0.197} & 0.390 & \textbf{0.112} & 0.396 & \textbf{0.112} \\
RecurBQR100   & 0.484 & \textbf{0.155} & 0.133 & \textbf{0.040} & 0.063 & \textbf{0.016} \\
RecurBQR1000  & 0.467 & \textbf{0.152} & 0.103 & \textbf{0.032} & 0.031 & \textbf{0.007} \\
RecurGLB10    & 0.246 & \textbf{0.090} & 0.152 & \textbf{0.052} & 0.152 & \textbf{0.053} \\
RecurGLB100   & 0.183 & \textbf{0.071} & 0.060 & \textbf{0.020} & 0.035 & \textbf{0.009} \\
RecurGLB1000  & 0.175 & \textbf{0.069} & 0.049 & \textbf{0.016} & 0.023 & \textbf{0.004} \\
RespondBQR10  & 1.144 & \textbf{0.288} & 0.816 & \textbf{0.199} & 0.814 & \textbf{0.198} \\
RespondBQR100 & 0.777 & \textbf{0.210} & 0.245 & \textbf{0.063} & 0.137 & \textbf{0.033} \\
RespondBQR1000& 0.718 & \textbf{0.196} & 0.156 & \textbf{0.042} & 0.044 & \textbf{0.010} \\
RespondGLB10  & 0.649 & \textbf{0.168} & 0.420 & \textbf{0.102} & 0.425 & \textbf{0.101} \\
RespondGLB100 & 0.449 & \textbf{0.127} & 0.133 & \textbf{0.035} & 0.070 & \textbf{0.016} \\
RespondGLB1000& 0.417 & \textbf{0.123} & 0.095 & \textbf{0.028} & 0.031 & \textbf{0.006} \\
\bottomrule
\end{tabular}
}
\end{table}

\begin{table}[htbp]
\centering
\footnotesize
\caption{\small Execution time (s): Reelay binary vs.\ LoomRV binary, single-property, dense-time.}
\label{tab:dense_binary}
\resizebox{\columnwidth}{!}{
\begin{tabular}{l rr rr rr}
\toprule
& \multicolumn{2}{c}{\textbf{Dense1 (s)}} & \multicolumn{2}{c}{\textbf{Dense10 (s)}} & \multicolumn{2}{c}{\textbf{Dense100 (s)}} \\
\cmidrule(r){2-3} \cmidrule(r){4-5} \cmidrule(r){6-7}
\textbf{Benchmark} & \textbf{Reelay} & \textbf{LoomRV} & \textbf{Reelay} & \textbf{LoomRV} & \textbf{Reelay} & \textbf{LoomRV} \\
\midrule
AbsentAQ10    & 0.378 & \textbf{0.084} & 0.220 & \textbf{0.048} & 0.209 & \textbf{0.046} \\
AbsentAQ100   & 0.332 & \textbf{0.077} & 0.169 & \textbf{0.039} & 0.143 & \textbf{0.034} \\
AbsentAQ1000  & 0.326 & \textbf{0.077} & 0.159 & \textbf{0.038} & 0.135 & \textbf{0.033} \\
AbsentBQR10   & 0.403 & \textbf{0.131} & 0.282 & \textbf{0.087} & 0.287 & \textbf{0.088} \\
AbsentBQR100  & 0.261 & \textbf{0.096} & 0.087 & \textbf{0.029} & 0.051 & \textbf{0.015} \\
AbsentBQR1000 & 0.239 & \textbf{0.091} & 0.060 & \textbf{0.021} & 0.023 & \textbf{0.006} \\
AbsentBR10    & 0.299 & \textbf{0.066} & 0.184 & \textbf{0.039} & 0.182 & \textbf{0.038} \\
AbsentBR100   & 0.277 & \textbf{0.063} & 0.161 & \textbf{0.035} & 0.142 & \textbf{0.032} \\
AbsentBR1000  & 0.274 & \textbf{0.063} & 0.155 & \textbf{0.035} & 0.141 & \textbf{0.031} \\
AlwaysAQ10    & 0.391 & \textbf{0.081} & 0.216 & \textbf{0.046} & 0.210 & \textbf{0.044} \\
AlwaysAQ100   & 0.350 & \textbf{0.075} & 0.163 & \textbf{0.038} & 0.133 & \textbf{0.032} \\
AlwaysAQ1000  & 0.343 & \textbf{0.074} & 0.157 & \textbf{0.037} & 0.127 & \textbf{0.031} \\
AlwaysBQR10   & 0.596 & \textbf{0.116} & 0.413 & \textbf{0.075} & 0.407 & \textbf{0.075} \\
AlwaysBQR100  & 0.402 & \textbf{0.084} & 0.126 & \textbf{0.026} & 0.067 & \textbf{0.012} \\
AlwaysBQR1000 & 0.369 & \textbf{0.079} & 0.084 & \textbf{0.018} & 0.028 & \textbf{0.005} \\
AlwaysBR10    & 0.310 & \textbf{0.063} & 0.183 & \textbf{0.037} & 0.177 & \textbf{0.035} \\
AlwaysBR100   & 0.289 & \textbf{0.060} & 0.154 & \textbf{0.033} & 0.133 & \textbf{0.029} \\
AlwaysBR1000  & 0.287 & \textbf{0.060} & 0.150 & \textbf{0.032} & 0.130 & \textbf{0.029} \\
RecurBQR10    & 0.545 & \textbf{0.133} & 0.329 & \textbf{0.078} & 0.333 & \textbf{0.077} \\
RecurBQR100   & 0.412 & \textbf{0.106} & 0.111 & \textbf{0.028} & 0.053 & \textbf{0.012} \\
RecurBQR1000  & 0.387 & \textbf{0.104} & 0.086 & \textbf{0.023} & 0.026 & \textbf{0.005} \\
RecurGLB10    & 0.210 & \textbf{0.057} & 0.131 & \textbf{0.033} & 0.131 & \textbf{0.033} \\
RecurGLB100   & 0.150 & \textbf{0.045} & 0.050 & \textbf{0.013} & 0.029 & \textbf{0.006} \\
RecurGLB1000  & 0.143 & \textbf{0.043} & 0.040 & \textbf{0.010} & 0.019 & \textbf{0.004} \\
RespondBQR10  & 0.949 & \textbf{0.201} & 0.693 & \textbf{0.140} & 0.679 & \textbf{0.141} \\
RespondBQR100 & 0.663 & \textbf{0.145} & 0.202 & \textbf{0.044} & 0.112 & \textbf{0.024} \\
RespondBQR1000& 0.607 & \textbf{0.136} & 0.129 & \textbf{0.030} & 0.036 & \textbf{0.008} \\
RespondGLB10  & 0.570 & \textbf{0.106} & 0.369 & \textbf{0.064} & 0.369 & \textbf{0.065} \\
RespondGLB100 & 0.380 & \textbf{0.081} & 0.112 & \textbf{0.022} & 0.059 & \textbf{0.010} \\
RespondGLB1000& 0.352 & \textbf{0.077} & 0.080 & \textbf{0.018} & 0.026 & \textbf{0.005} \\
\bottomrule
\end{tabular}
}
\end{table}

\subsection{Multi-Property Monitoring Performance}
\label{sec:eval-multi}
This experiment evaluates the end-to-end throughput of our multi-property framework when all 30 \textit{timescales} benchmark properties~\cite{ulus2019timescales} are monitored simultaneously.

\paragraph{Configurations}
We compare five configurations that progressively isolate the contributions of the linearized execution schedule, the zero-allocation arena, and the shared DAG:
\begin{itemize}
  \item \textbf{Reelay-Sequential}: 30 independent single-property Reelay monitors, each processing the full trace in a separate pass.
  \item \textbf{Reelay-AND}: all 30 properties conjoined into a single formula via explicit conjunction operators ($\wedge$), monitored by one Reelay instance. This simulates a best-case single-monitor strategy without structural deduplication.
  \item \textbf{LoomRV Sequential}: 30 independent LoomRV monitors, one per property. This isolates the benefit of the linearized execution schedule and zero-allocation arena described in Sections~\ref{sec:frontend} and~\ref{sec:runtime}, without any cross-property sharing.
  \item \textbf{LoomRV-AND}: all 30 properties conjoined into a single formula and compiled by LoomRV. Because the content-addressable node database (Section~\ref{sec:frontend}) deduplicates shared subformulas even within a single conjunction, this configuration benefits from both the data-oriented execution model and partial structural reuse, but retains the 29 extra conjunction nodes.
  \item \textbf{LoomRV Multi}: the full multi-property monitor, compiling all 30 properties into a single shared DAG with one root node per property. This is the primary configuration of this work.
\end{itemize}
LoomRV is evaluated in both JSON and binary feeder modes in every configuration.

\begin{table}[htbp]
    \centering
    \caption{Total wall-clock time (s) to monitor 30 \PastMTL\ properties over 1\,000\,000 discrete-time steps.
    Speedup is relative to the Reelay-Sequential baseline for the same feeder type.}
    \label{tab:multi-property-discrete}
    \begin{tabular}{l r r}
        \toprule
        \textbf{Configuration} & \textbf{Time (s)} & \textbf{Speedup} \\
        \midrule
        \multicolumn{3}{l}{\textit{JSON feeder}} \\
        \midrule
        Reelay-Sequential (JSON)             & 9.34 & $1.0\times$ (baseline) \\
        Reelay-AND (JSON)                    & 6.38 & $1.5\times$ \\
        LoomRV Sequential (JSON) & 4.42 & $2.1\times$ \\
        LoomRV-AND (JSON)        & 0.80 & $11.6\times$ \\
        LoomRV Multi (JSON)      & \textbf{0.80} & $\mathbf{11.7\times}$ \\
        \midrule
        \multicolumn{3}{l}{\textit{Binary feeder}} \\
        \midrule
        Reelay-Sequential (binary)             & 4.66 & $1.0\times$ (baseline) \\
        Reelay-AND (binary)                    & 3.63 & $1.3\times$ \\
        LoomRV Sequential (binary) & 1.58 & $3.0\times$ \\
        LoomRV-AND (binary)        & \textbf{0.70} & $\mathbf{6.6\times}$ \\
        LoomRV Multi (binary)      & 0.71 & $6.6\times$ \\

        \bottomrule
    \end{tabular}
\end{table}

\begin{table}[htbp]
    \centering
    \caption{Total wall-clock time (s) to monitor 30 \PastMTL\ properties over 1\,000\,000 dense-time steps.
    Speedup is relative to the Reelay-Sequential baseline for the same feeder type.}
    \label{tab:multi-property}
    \begin{tabular}{l r r}
        \toprule
        \textbf{Configuration} & \textbf{Time (s)} & \textbf{Speedup} \\
        \midrule
                \multicolumn{3}{l}{\textit{JSON feeder}} \\
        \midrule
        Reelay-Sequential (JSON)             & 27.49 & $1.0\times$ (baseline) \\
        Reelay-AND (JSON)                    & 24.26 & $1.1\times$ \\
        LoomRV Sequential (JSON) & 9.63 & $2.9\times$ \\
        LoomRV-AND (JSON)        &  4.05 & $6.8\times$ \\
        LoomRV Multi (JSON)      &  \textbf{3.77} & $\mathbf{7.3\times}$ \\
        \midrule
        \multicolumn{3}{l}{\textit{Binary feeder}} \\
        \midrule
        Reelay-Sequential (binary)             & 22.28 & $1.0\times$ (baseline) \\
        Reelay-AND (binary)                    & 21.78 & $1.0\times$ \\
        LoomRV Sequential (binary) &  5.80 & $3.8\times$ \\
        LoomRV-AND (binary)        &  3.72 & $6.0\times$ \\
        LoomRV Multi (binary)      &  \textbf{3.45} & $\mathbf{6.4\times}$ \\

        \bottomrule
    \end{tabular}
\end{table}

Tables~\ref{tab:multi-property-discrete} and~\ref{tab:multi-property} report the results for discrete and dense time, respectively.

\paragraph{Execution model contribution}
Even without cross-property sharing, LoomRV Sequential already achieves $2.1$--$3.0\times$ speedup over Reelay-Sequential in discrete-time and $2.9$--$3.8\times$ in dense-time (lower and upper bounds correspond to JSON and binary feeders, respectively). This improvement is consistent with the benefits of the linearized execution schedule and the zero-allocation arena, confirming that the data-oriented execution model contributes independently of shared computation.

\paragraph{Effect of structural deduplication}
Reelay-AND, which eliminates redundant trace parsing by conjoining all properties into a single formula, yields only a modest $1.0$--$1.5\times$ improvement over Reelay-Sequential, since its object-oriented evaluator still processes every node independently.
In contrast, both LoomRV-AND and LoomRV Multi combine single-pass evaluation with the content-addressable deduplication of Section~\ref{sec:frontend}, achieving $6.0$--$11.7\times$ speedup over Reelay-Sequential across both time models and feeders.

\paragraph{Multi vs.\ AND conjunction}
In discrete-time, LoomRV Multi and LoomRV-AND perform nearly identically ($0.80$ vs $0.80$\,s JSON; $0.71$ vs $0.70$\,s binary). Because discrete-time node evaluation operates on scalar Booleans, the cost of the 29 additional conjunction nodes present in the AND formulation is negligible.
In dense-time, however, each node evaluation involves interval-set operations, making those extra nodes measurable: the shared DAG contains 107 nodes compared to 136 for the AND formulation. Accordingly, LoomRV Multi achieves the best overall results ($7.3\times$ JSON, $6.4\times$ binary), outperforming LoomRV-AND by a margin that directly reflects the avoided conjunction nodes ($3.77$ vs $4.05$\,s; $3.45$ vs $3.72$\,s).

\section{Related Work}
\label{sec:related}
The idea of executing temporal logic specifications dates back to the seminal work on Tempura~\cite{moszkowski1984executing} and the Metatem framework~\cite{gabbay1989declarative}. The latter specifically leverages past-time reasoning to enable monitoring and reactive execution. A closely related paradigm appears in synchronous languages such as Lustre~\cite{halbwachs2002synchronous}, Signal~\cite{gautier1987signal}, and Esterel~\cite{boussinot1991esterel}, where reactive programs are compiled into deterministic code under a global logical clock. Our work follows this synchronous execution model and focuses on the unified evaluation of multiple past temporal logic specifications.

Modern runtime verification frameworks build on these foundations to monitor temporal properties over execution traces. Early work~\cite{havelund2004monitoring} influenced the development of modern tools~\cite{pike2010copilot, johannsen2023r2u2} as well as frameworks with extended semantics~\cite{monpoly2, dejavu, ulus2026reelay}. Another line of work is stream-based monitoring~\cite{lola,gorostiaga2018striver,kallwies2022tessla}, closely resembling synchronous programming languages in both syntax and execution semantics. In these approaches, each specification is typically compiled into an independent monitor evaluated over the same trace. Despite the practical importance of monitoring large sets of properties, the problem of multi-property monitoring has received comparatively limited attention. Existing efforts addressing multiple properties primarily arise in formal verification contexts~\cite{goldberg2018efficient, dureja2019boosting, das2024purse, das2025sisco, roy2026mpbmc}, where the focus is on state-space reduction and incremental evaluation. 

In contrast, our work targets the online runtime verification setting and treats multi-property monitoring as a first-class citizen, combining cross-property subformula deduplication with a data-oriented execution model. Related ideas appear in optimizations for runtime monitors based on common subexpression elimination~\cite{schwenger2020automatic, johannsen2023r2u2, baumeister2025intermediate}; our approach generalizes this principle from local expression-level optimization to a multi-property construction integrated with the execution model.

\section{Conclusion}
\label{sec:conclusion}

In this paper, we presented \emph{LoomRV}, a highly efficient runtime monitoring framework designed from the ground up for the simultaneous evaluation of multiple \PastMTL\ properties. By compiling specifications into a deduplicated, linearized execution schedule and employing a double-buffered arena for zero-allocation state management, \emph{LoomRV} significantly mitigates the redundant computation and memory allocation overhead that traditionally limits scaling to large specification sets. Our comprehensive evaluation demonstrates per-property throughput improvements of $2\times$ to $4.5\times$ from the data-oriented execution model alone, scaling to $6\times$ to $12\times$ end-to-end when monitoring 30 properties simultaneously, outperforming the state-of-the-art sequential network monitor \emph{Reelay} across both discrete and dense-time traces.

Our findings also highlight a pressing need for standardized multi-property benchmark suites, since existing benchmarks target single-property evaluation and the degree of subexpression overlap in industrial specification sets remains difficult to quantify.

Future extensions will focus on deployment over publish-subscribe networks for distributed monitoring, and on FPGA-based hardware synthesis, for which the zero-allocation, contiguous memory layout is inherently well suited.

\bibliographystyle{IEEEtran}
\bibliography{references}

\begin{thebibliography}{10}
\providecommand{\url}[1]{#1}
\csname url@samestyle\endcsname
\providecommand{\newblock}{\relax}
\providecommand{\bibinfo}[2]{#2}
\providecommand{\BIBentrySTDinterwordspacing}{\spaceskip=0pt\relax}
\providecommand{\BIBentryALTinterwordstretchfactor}{4}
\providecommand{\BIBentryALTinterwordspacing}{\spaceskip=\fontdimen2\font plus
\BIBentryALTinterwordstretchfactor\fontdimen3\font minus \fontdimen4\font\relax}
\providecommand{\BIBforeignlanguage}[2]{{%
\expandafter\ifx\csname l@#1\endcsname\relax
\typeout{** WARNING: IEEEtran.bst: No hyphenation pattern has been}%
\typeout{** loaded for the language `#1'. Using the pattern for}%
\typeout{** the default language instead.}%
\else
\language=\csname l@#1\endcsname
\fi
#2}}
\providecommand{\BIBdecl}{\relax}
\BIBdecl

\bibitem{pnueli1977ltl}
A.~Pnueli, ``The temporal logic of programs,'' in \emph{Proceedings of the Symposium on Foundations of Computer Science (FOCS)}, 1977, pp. 46--57.

\bibitem{koymans1990mtl}
R.~Koymans, ``Specifying real-time properties with metric temporal logic,'' \emph{Real-Time Systems}, vol.~2, no.~4, pp. 255--299, 1990.

\bibitem{ulus2026online}
D.~Ulus, ``Online monitoring of metric temporal logic using sequential networks,'' \emph{Logical Methods in Computer Science}, vol.~22, 2026.

\bibitem{ulus2026reelay}
------, ``Reelay: Online temporal logic monitoring framework,'' \emph{arXiv preprint arXiv:2604.22384}, 2026.

\bibitem{langdale2019parsing}
G.~Langdale and D.~Lemire, ``Parsing gigabytes of {JSON} per second,'' \emph{The VLDB Journal}, vol.~28, no.~6, pp. 941--960, 2019.

\bibitem{ulus2019timescales}
D.~Ulus, ``Timescales: A benchmark generator for {MTL} monitoring tools,'' in \emph{International Conference on Runtime Verification}.\hskip 1em plus 0.5em minus 0.4em\relax Springer, 2019, pp. 402--412.

\bibitem{moszkowski1984executing}
B.~Moszkowski, ``Executing temporal logic programs: preliminary version,'' in \emph{International Conference on Concurrency}.\hskip 1em plus 0.5em minus 0.4em\relax Springer, 1984, pp. 111--130.

\bibitem{gabbay1989declarative}
D.~Gabbay, ``The declarative past and imperative future: Executable temporal logic for interactive systems,'' in \emph{Temporal Logic in Specification}, ser. LNCS, vol. 398.\hskip 1em plus 0.5em minus 0.4em\relax Springer, 1989, pp. 409--448.

\bibitem{halbwachs2002synchronous}
N.~Halbwachs, P.~Caspi, P.~Raymond, and D.~Pilaud, ``The synchronous data flow programming language lustre,'' \emph{Proceedings of the IEEE}, vol.~79, no.~9, pp. 1305--1320, 1991.

\bibitem{gautier1987signal}
T.~Gautier, P.~Le~Guernic, and L.~Besnard, ``Signal: A declarative language for synchronous programming of real-time systems,'' in \emph{Conference on Functional Programming Languages and Computer Architecture}.\hskip 1em plus 0.5em minus 0.4em\relax Springer, 1987, pp. 257--277.

\bibitem{boussinot1991esterel}
F.~Boussinot and R.~De~Simone, ``The esterel language,'' \emph{Proceedings of the IEEE}, vol.~79, no.~9, pp. 1293--1304, 1991.

\bibitem{havelund2004monitoring}
K.~Havelund and G.~Ro{\c{s}}u, ``Efficient monitoring of safety properties,'' \emph{International Journal on Software Tools for Technology Transfer}, vol.~6, no.~2, pp. 158--173, 2004.

\bibitem{pike2010copilot}
L.~Pike, A.~Goodloe, R.~Morisset, and S.~Niller, ``Copilot: A hard real-time runtime monitor,'' in \emph{International Conference on Runtime Verification}.\hskip 1em plus 0.5em minus 0.4em\relax Springer, 2010, pp. 345--359.

\bibitem{johannsen2023r2u2}
C.~Johannsen, P.~Jones, B.~Kempa, K.~Y. Rozier, and P.~Zhang, ``R2u2 version 3.0: Re-imagining a toolchain for specification, resource estimation, and optimized observer generation for runtime verification in hardware and software,'' in \emph{International Conference on Computer Aided Verification}.\hskip 1em plus 0.5em minus 0.4em\relax Springer, 2023, pp. 483--497.

\bibitem{monpoly2}
D.~Basin, F.~Klaedtke, and E.~Zalinescu, ``The {MonPoly} monitoring tool,'' in \emph{RV-CuBES 2017. An International Workshop on Competitions, Usability, Benchmarks, Evaluation, and Standardisation for Runtime Verification Tools}, vol.~3, 2017, pp. 19--28.

\bibitem{dejavu}
K.~Havelund, D.~Peled, and D.~Ulus, ``Dejavu: A monitoring tool for first-order temporal logic,'' in \emph{Proceedings of the Workshop on Monitoring and Testing of Cyber-Physical Systems (MT-CPS)}, 2018, pp. 12--13.

\bibitem{lola}
B.~D'Angelo, S.~Sankaranarayanan, C.~S{\'a}nchez, W.~Robinson, B.~Finkbeiner, H.~B. Sipma, S.~Mehrotra, and Z.~Manna, ``Lola: Runtime monitoring of synchronous systems,'' in \emph{Proceedings of the Symposium on Temporal Representation and Reasoning (TIME)}, 2005, pp. 166--174.

\bibitem{gorostiaga2018striver}
F.~Gorostiaga and C.~S{\'a}nchez, ``Striver: Stream runtime verification for real-time event-streams,'' in \emph{International Conference on Runtime Verification}.\hskip 1em plus 0.5em minus 0.4em\relax Springer, 2018, pp. 282--298.

\bibitem{kallwies2022tessla}
H.~Kallwies, M.~Leucker, M.~Schmitz, A.~Schulz, D.~Thoma, and A.~Weiss, ``Tessla--an ecosystem for runtime verification,'' in \emph{International Conference on Runtime Verification}.\hskip 1em plus 0.5em minus 0.4em\relax Springer, 2022, pp. 314--324.

\bibitem{goldberg2018efficient}
E.~Goldberg, M.~G{\"u}demann, D.~Kroening, and R.~Mukherjee, ``Efficient verification of multi-property designs (the benefit of wrong assumptions),'' in \emph{2018 Design, Automation \& Test in Europe Conference \& Exhibition (DATE)}.\hskip 1em plus 0.5em minus 0.4em\relax IEEE, 2018, pp. 43--48.

\bibitem{dureja2019boosting}
R.~Dureja, J.~Baumgartner, A.~Ivrii, R.~Kanzelman, and K.~Y. Rozier, ``Boosting verification scalability via structural grouping and semantic partitioning of properties,'' in \emph{2019 Formal Methods in Computer Aided Design (FMCAD)}.\hskip 1em plus 0.5em minus 0.4em\relax IEEE, 2019, pp. 1--9.

\bibitem{das2024purse}
S.~Das, A.~Hazra, P.~Dasgupta, S.~Kundu, and H.~Jain, ``{PURSE}: {P}roperty ordering using runtime statistics for efficient multi-property verification,'' in \emph{2024 Design, Automation \& Test in Europe Conference \& Exhibition (DATE)}.\hskip 1em plus 0.5em minus 0.4em\relax IEEE, 2024, pp. 1--6.

\bibitem{das2025sisco}
S.~Das, A.~Hazra, P.~Dasgupta, H.~Jain, and S.~Kundu, ``Sisco: Selective invariant sharing, clustering and ordering for effective multi-property formal verification,'' in \emph{Proceedings of the 30th Asia and South Pacific Design Automation Conference}, 2025, pp. 1343--1349.

\bibitem{roy2026mpbmc}
S.~G. Roy, S.~Ghosh, A.~Banerjee, R.~K. Gajavelly, and S.~Surendran, ``Mpbmc: Multi-property bounded model checking with gnn-guided clustering,'' in \emph{2026 39th International Conference on VLSI Design \& 25th International Conference on Embedded Systems (VLSID)}.\hskip 1em plus 0.5em minus 0.4em\relax IEEE, 2026, pp. 179--184.

\bibitem{schwenger2020automatic}
J.~Baumeister, B.~Finkbeiner, M.~Kruse, and M.~Schwenger, ``Automatic optimizations for stream-based monitoring languages,'' in \emph{International Conference on Runtime Verification}.\hskip 1em plus 0.5em minus 0.4em\relax Springer, 2020, pp. 451--461.

\bibitem{baumeister2025intermediate}
J.~Baumeister, A.~Correnson, B.~Finkbeiner, and F.~Scheerer, ``An intermediate program representation for optimizing stream-based languages,'' in \emph{International Conference on Computer Aided Verification}.\hskip 1em plus 0.5em minus 0.4em\relax Springer, 2025, pp. 393--407.

\end{thebibliography}

\end{document}